\documentclass[useAMS,usenatbib]{mn2e} 
\pdfpageheight=\paperheight
\pdfpagewidth=\paperwidth
\usepackage{graphicx}
\usepackage{rotating}
\usepackage[colorlinks=true, linkcolor=blue, citecolor=blue, urlcolor=blue]{hyperref}

\title[Weak turbulence in the core of Abell 1413]
{Weak cool core, weak turbulence: \textit{XRISM}, \textit{XMM--Newton} and \textit{Chandra}
spectroscopy of the core of Abell 1413}
\author[E. G\"ormez and M. Hudaverdi]{E. G\"ormez$^{1,2}$ and M. Hudaverdi$^{1}$\thanks{E-mail: hudaverd@yildiz.edu.tr}\\
$^{1}$Department of Physics, Faculty of Science and Art, Y{\i}ld{\i}z Technical University, 34220, Istanbul, Turkey\\
$^{2}$Graduate School of Natural and Applied Sciences, Y{\i}ld{\i}z  Technical University, Istanbul,Turkey\\
}

\begin{document}

\date{Accepted ------------------ . Received ------------------ ; in original form ------------------}

\pagerange{\pageref{firstpage}--\pageref{lastpage}}\pubyear{2002}

\maketitle
\label{firstpage}
\begin{abstract}
We present the first \textit{XRISM} spectroscopy of the core of Abell~1413 ($z = 0.143$), 
combining $104.7$~ks of Resolve with archival \textit{XMM--Newton} and \textit{Chandra} spectra of a matched sky region.
A1413 is hot, massive and relaxed but lacks a strong cool core ($t_{\rm cool} = 6.4$~Gyr), 
a regime in which the low turbulence reported to date has been tested only once, in A3571. 
The Resolve spectrum requires a single velocity-broadened component with $kT = 7.06^{+0.33}_{-0.43}$~keV and
a line-of-sight velocity dispersion $\sigma_{v} = 170^{+25}_{-24}$~km~s$^{-1}$ ($\mathcal{M}_{\rm 3D} = 0.217$). 
The bulk velocity relative to the brightest cluster galaxy, $v_{\rm bulk} = -24 \pm 26\,(z_{\rm ICM}) \pm 71\,(z_{\rm BCG})$~km~s$^{-1}$, 
is limited by the optical redshift rather than by Resolve, and disfavours coherent line-of-sight sloshing.
The non-thermal pressure fraction, $2.5 \pm 0.7$~per cent, implies a hydrostatic mass bias $b \la 2.5$~per cent, 
comparable to the strong cool-core cluster A2029 ($2.6\pm0.3$~per cent) and, at matched sampling scale, if anything below it. 
With A3571 ($1.3\pm0.3$~per cent), to which A1413 rescales within $0.2\sigma$, 
two weak cool cores now show that this quiescence does not require central cooling. 
The resolved Fe--K He$\alpha$ complex gives Fe~$= 0.38 \pm 0.03\,({\rm stat}) \pm 0.04\,({\rm sys})\,Z_\odot$; 
Ni/Fe, separated from Fe~K$\beta$ for the first time, is solar, as is the pattern as a whole. 
Ne/Fe, Mg/Fe and Si/Fe lie below $2$~keV, where our instruments disagree by $10$--$20$~per cent; 
we therefore do not regard the SNIa fraction as constrained.
\end{abstract}


\section{Introduction}
\label{sec:intro}
 
Galaxy clusters are the largest virialized structures in the Universe, and the properties of their hot, X-ray-emitting intracluster medium (ICM) provide some of the most powerful tools for constraining structure formation and cosmological parameters \citep{Allen2011,Pratt2019}. 
A large fraction of clusters host a cool core: a dense, 
low-entropy central region where the radiative cooling time falls well below the Hubble time \citep{Molendi2001,Hudson2010}. 
Left unchecked, this cooling should drive a cooling flow of hundreds to thousands of solar masses per year onto the central galaxy, 
yet such flows are not observed at the predicted rate \citep{Peterson2006}. 
This ``cooling flow problem'' is now understood to be resolved by mechanical feedback from the central active galactic nucleus (AGN), 
which injects energy into the surrounding gas through jets, cavities, shocks and turbulence, offsetting radiative losses in a quasi-self-regulated cycle
\citep{McNamara2012,Gaspari2012}. 
This feedback loop does not merely heat the gas thermally: it also stirs the ICM, seeding it with bulk flows and turbulence 
whose dynamical importance and dissipation timescale remain poorly constrained observationally \citep{Hitomi2016}. 
Because the standard method for deriving cluster masses from X-ray data assumes hydrostatic equilibrium (HSE), 
any unaccounted-for non-thermal pressure support --- from turbulence, bulk motions, cosmic rays or magnetic fields --- 
causes HSE-based masses to be biased low relative to the true gravitating mass, 
an effect known as the hydrostatic mass bias \citep{Nagai2007,Nelson2014}. 
This bias propagates directly into cluster mass function analyses and, ultimately, into cosmological parameter estimates derived from cluster
abundance studies \citep{Pratt2019}, making its accurate quantification --- particularly in cluster cores, where feedback-driven motions are expected to be strongest --- a problem of both astrophysical and cosmological significance.

Directly measuring this non-thermal pressure component, however, has long been beyond the reach of CCD-resolution X-ray instruments.
Detectors such as \textit{XMM--Newton}/EPIC and \textit{Chandra}/ACIS, with energy resolutions of order 50--150~eV at the Fe--K complex, 
cannot resolve the line-of-sight velocity shifts ($\sim$10--100 km s$^{-1}$) or the velocity broadening produced by turbulence, 
since these correspond to Doppler shifts and line widths far narrower than the instrumental resolution. 
Non-thermal pressure fractions have therefore historically been inferred only indirectly --- from surface-brightness fluctuations \citep{Zhuravleva2014},
from resonant scattering of the brightest resonance lines \citep{Sanders2006}, 
or by comparison with hydrodynamical simulations \citep{Lau2009,Angelinelli2020} --- each carrying substantial model-dependent uncertainty. Compounding this limitation, the same low spectral resolution leaves elemental abundance measurements degenerate 
with the assumed temperature structure of the gas: 
multi-temperature and single-temperature fits to CCD-resolution spectra can yield significantly different abundances for the same underlying data,
particularly for the Fe-peak and $\alpha$-elements that trace the relative enrichment by Type~Ia and 
core-collapse supernovae \citep{dePlaa2007,Mernier2018}. 
Without an instrument capable of resolving individual emission lines, 
both the dynamical state and the detailed chemical enrichment history of cluster cores have remained only loosely constrained.

The X-Ray Imaging and Spectroscopy Mission (\textit{XRISM}), launched in 2023, addresses both limitations directly through its Resolve
microcalorimeter, which achieves $\sim5$~eV FWHM energy resolution across the 1.7--12~keV band \citep{Tashiro2020,Ishisaki2022}. 
At this resolution, Resolve measures line-of-sight bulk velocities and turbulent velocity dispersions from the centroids 
and widths of resolved emission lines to an accuracy of tens of km~s$^{-1}$, 
providing direct spectroscopic constraints on ICM gas motions \citep{Hitomi2018,XRISM2025a}. 
The same resolution breaks the temperature--abundance degeneracies that limit CCD spectroscopy,
enabling simultaneous constraints on multiple elemental abundances even in the presence of complex, multi-temperature gas. 
The first \textit{XRISM} measurements of cluster cores have consistently revealed quiescent gas: line-of-sight velocity dispersions of
$100$--$200$~km~s$^{-1}$ and non-thermal pressure fractions of only a few \% in A2029 \citep{XRISM2025a}, A1795 \citep{Sarkar2026a1795} 
and the Centaurus cluster \citep{XRISM2025b}, in tension with $\approx$10\% predicted by cosmological simulations \citep{Nelson2014}. 
Crucially, however, every cluster core measured to date is a strong cool core.
Whether the low levels of turbulence found so far are a general property of relaxed clusters, or a consequence of this selection, remains untested.

Abell~1413 ($z = 0.143$) offers a direct test of that question. 
It is a massive, hot ($kT \sim 7$--$8$~keV) cluster long classified as dynamically relaxed on the basis of its smooth, circularly symmetric
X-ray surface brightness and the good agreement between its X-ray and Sunyaev--Zel'dovich centroids \citep{Allen1998,AMI2012}. 
Deep \textit{Chandra} imaging shows no clear evidence of a recent major merger, 
aside from a weak surface-brightness edge $\sim400$~kpc north of the cluster centre \citep{Vikhlinin2005}, 
and its temperature and density profiles derived independently with \textit{XMM--Newton},
\textit{Chandra} and \textit{Suzaku} agree well out to half the virial radius and beyond \citep{Pratt2002,Bonamente2006,Hoshino2010}. 
Yet A1413 does not display the pronounced central temperature drop that typically defines a cool core in relaxed clusters of comparable mass:
only the innermost radial bin shows a modest decline, and some studies suggest that even this may be partly a consequence of PSF blurring
rather than a genuine cool core \citep{Pratt2002,Vikhlinin2006}. 
A weak central radio source and a tentative, offset mini-halo have also been reported in the core \citep{Govoni2009, Lusetti2024}, 
hinting at some level of recent AGN or gas-dynamical activity that CCD-resolution spectroscopy has been unable to characterise kinematically.

\begin{table*}
\caption{Summary of the A1413 observations analysed in this paper.\label{tab:obslog}}   
\begin{center}
\setlength{\tabcolsep}{3pt}
\begin{tabular}{ccccccccc}
\hline
\hline
Telescope 	& Instrument		& Obs. ID 		& RA 		& Dec. 		& Obs. Date 		& Total Exp. 	& Net Exp. \\
 			& 				& 			&  			&  			& 				& (ks) 		&  (ks)\\
\hline	
\hline
Chandra 		& ACIS-I 			&  5003 		& 11$^{h}$55$^{m}$18$^{s}$.10 & +23$^\circ$24$^\prime$17.0$^{\prime\prime}$ 	& 2004-03-06 		& 76.05 			& 75.05	\\
			& ACIS-I 			&  5002 		& 11$^{h}$55$^{m}$09$^{s}$.60 & +23$^\circ$30$^\prime$46.8$^{\prime\prime}$ 	& 2005-02-03 		& 37.15 			& 36.67 	\\
\hline
XMM-Newton 	& PN/MOS/RGS 	&  0502690101 & 11$^{h}$55$^{m}$18$^{s}$.91 & +23$^\circ$24$^\prime$13.8$^{\prime\prime}$ 	& 2007-11-25 		& 66.65 			& 51.15/57.73/59.17 \\
			& PN/MOS/RGS 	&  0502690201 & 11$^{h}$55$^{m}$18$^{s}$.91 & +23$^\circ$24$^\prime$13.8$^{\prime\prime}$ 	& 2007-12-11 		& 82.45 			& 72.97/66.18/68.13    \\
			& PN/MOS/RGS 	&  0551280101 & 11$^{h}$55$^{m}$18$^{s}$.91 & +23$^\circ$24$^\prime$13.8$^{\prime\prime}$ 	& 2008-12-04 		& 76.72 			& 61.55/61.21/65.06\\
			& PN/MOS/RGS	&  0551280201 & 11$^{h}$55$^{m}$18$^{s}$.91 & +23$^\circ$24$^\prime$13.8$^{\prime\prime}$	& 2008-12-06  		& 76.15  			& 59.17/64.50/63.53 \\
\hline
XRISM 		& Resolve/Xtend 	& 201049010  	& 11$^{h}$55$^{m}$18$^{s}$.55 & +23$^\circ$24$^\prime$37.4$^{\prime\prime}$ 	& 2024-11-15 		&107.82 			& 104.65/98.22 \\
\hline
\end{tabular}
\end{center}
\end{table*}

\begin{figure*}
\centering
  \vspace*{17pt}
 \includegraphics[width=17cm]{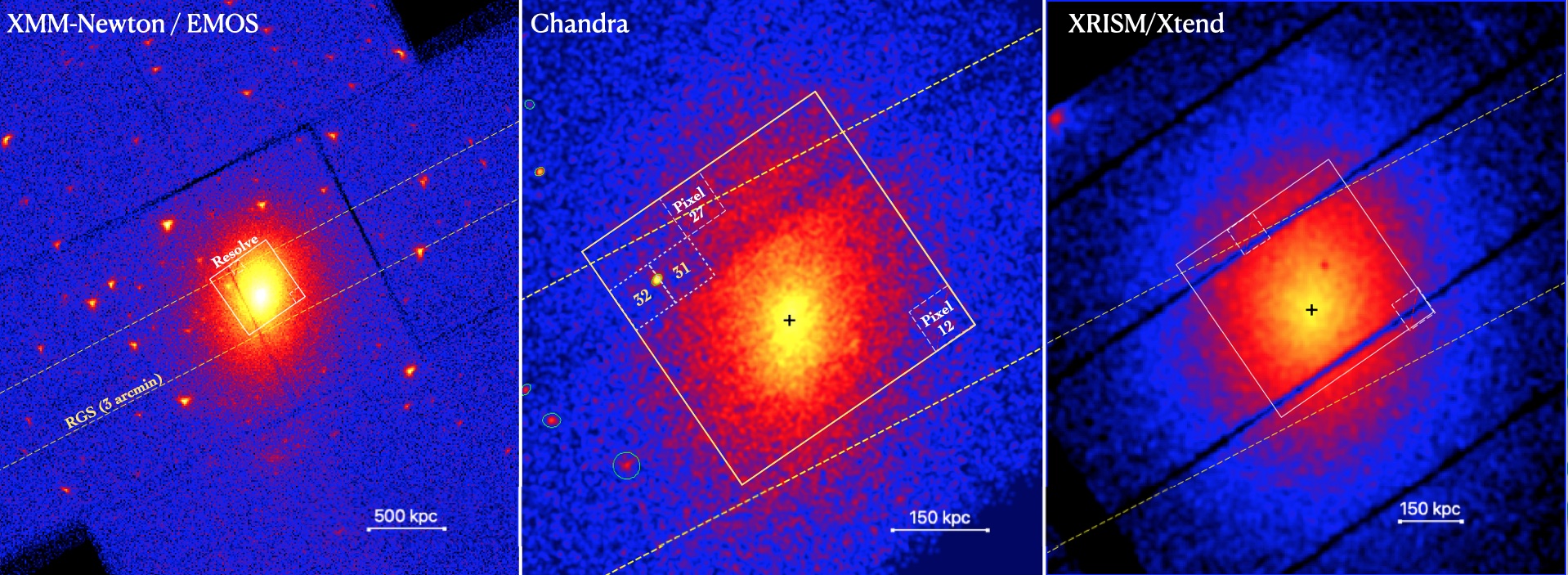}
  \caption{Exposure-corrected \textit{XMM-Newton}/MOS mosaic image of   A1413 in the 0.3--10\,keV band (left), 
  \textit{Chandra} 0.5--7\,keV merged image (centre), and \textit{XRISM}/Xtend image in the 0.5--10\,keV band (right). 
  The white box in each panel illustrates the \textit{XRISM}/Resolve field of view (FoV) (excluding pixels 12 and 27), 
  and the yellow cross-dispersion lines mark the RGS extraction regions.
  Pixels 31 and 32, adjacent to a point source, are also excluded from the Resolve spectral extraction. 
  All images have been adaptively smoothed and logarithmically scaled to highlight the structure of the extended plasma emission.}
     \label{fig:a1413images}
\end{figure*}

With a central cooling time of $t_{\rm cool} = 6.4 \pm 1.3$~Gyr within $r < 0.5'$ ($\sim$75~kpc), 
measured from the \textit{XMM--Newton} density and temperature profiles by \citet{Lusetti2024}, 
A1413 falls in the weak cool-core regime of \citet{Hudson2010}  (strong cool cores $t_{\rm cool}<1$~Gyr, weak $1$--$7.7$~Gyr, non-cool-core $>7.7$~Gyr).
It therefore occupies the same intermediate regime as A3571 \citep{McCall2026} --- morphologically relaxed, but lacking the strong
central cooling that characterises the other relaxed cluster cores observed with a microcalorimeter to date --- at roughly four times the redshift.
The absence of a pronounced central temperature drop may indicate limited feedback activity and a comparatively undisturbed core, 
or it may mask non-thermal pressure support and multi-temperature structure that CCD-resolution instruments cannot resolve --- 
a degeneracy that only velocity-resolved spectroscopy can break.
Its elemental abundance pattern, and by extension its enrichment history,
has likewise never been constrained at a resolution sufficient to separate the blended lines inherent to CCD detectors.

The measurement also probes a physical scale the microcalorimeter sample has not previously reached. 
At $z = 0.143$ the $3\arcmin\times3\arcmin$ Resolve array subtends $452$~kpc, against $138$~kpc for A3571 and $263$~kpc for A2029 
--- the largest region sampled in any relaxed cluster core observed with a microcalorimeter to date. 
Because turbulent amplitudes grow with the scale over which they are averaged, 
A1413 therefore tests whether the low dispersions reported in nearby cores persist when the ICM is sampled over a substantially larger volume, 
rather than simply repeating those measurements in a new object.

In this paper we present the first \textit{XRISM} observation of the core of A1413, combining Resolve and Xtend spectroscopy 
with archival \textit{XMM--Newton} (EPIC and RGS) and \textit{Chandra} (ACIS) data extracted from a matched sky region. 
We derive single-temperature ({\sc vapec}) and velocity-broadened ({\sc bvapec}) fits across all five datasets 
to obtain a self-consistent measurement of the temperature and elemental abundance structure of the core, 
and to cross-check for systematic offsets between instruments. 
From the Resolve line centroids and widths we measure the bulk and turbulent line-of-sight velocities of the ICM for the first time in this cluster, 
and derive the resulting non-thermal pressure fraction and its contribution to the hydrostatic mass bias. 
We further use the combined abundance pattern to test whether the enrichment of the core departs from solar element ratios, 
and discuss the result in the context of the cluster's revised dynamical picture.

The paper is organised as follows. 
Section~\ref{sec:obs} describes the observations and data reduction; 
Section~\ref{sec:spectral} presents the spectral modelling methodology; 
Section~\ref{sec:results} gives the resulting temperatures and elemental abundances and discusses the chemical enrichment of A1413; 
Section~\ref{sec:kinematics} derives the gas kinematics and the implications for the non-thermal pressure support and the hydrostatic mass bias; and
Section~\ref{sec:conclusions} summarises our conclusions.
Appendix~\ref{app:pointsources} assesses the effect of the point source projected onto Resolve and defines the pixel configuration adopted throughout.
Throughout we assume a flat $\Lambda$CDM cosmology with $H_0 = 70$~km~s$^{-1}$~Mpc$^{-1}$, $\Omega_{\rm m} = 0.3$ and 
$\Omega_\Lambda = 0.7$, consistent with recent \textit{XRISM} ICM kinematics papers \citep{XRISM2025a,Sarkar2026a1795}. 
At the redshift of A1413, $1'$ corresponds to approximately 151 kpc.

\section{Observations and data reduction}
\label{sec:obs}
 
The galaxy cluster A1413 has been the target of multiple X-ray observing campaigns across different space observatories. 
To compensate for the lack of \textit{XRISM}/Resolve spectral sensitivity below 2.0\,keV caused by the closed gate valve, 
we integrate the archival \textit{XMM-Newton} (EPIC and RGS) and \textit{Chandra} (ACIS) datasets ($>30$\,ks) into the analysis. 
A1413 was also observed with \textit{Suzaku}, 
pointed at the cluster outskirts ($R_{200}$), offset by 9.3\,arcmin from the cluster core \citep{Hoshino2010};
since this observation fails to cover the core region relevant to our analysis, we exclude it from our final dataset. 
The complete log of the observations analysed in this study is given in Table~\ref{tab:obslog}, 
and the three-instrument images of the cluster core are shown in Fig.~\ref{fig:a1413images}.
 
\subsection{XRISM/Resolve \& Xtend} 
\label{subsec:xrism}

\textit{XRISM} targeted the central core of A1413 (ObsID 201049010) over the period 2024 November 15--17, 
accumulating a total raw exposure of 107.82\,ks. 
The observation was carried out with both onboard instruments: 
Resolve, a $6\times6$ pixel microcalorimeter array providing non-dispersive spectroscopy 
with $\sim$5--7\,eV resolution across the Fe-K band \citep{Ishisaki2022}, 
and Xtend, a wide-field ($38\arcmin\times38\arcmin$) CCD imager covering 0.4--13\,keV \citep{Tashiro2020}. 
Data from both instruments were reprocessed using \textsc{HEASoft} v6.36 and \textit{XRISM} CalDB version 20250915.
 
Throughout Cycle 1, the inability to open the gate valve directly impacted the instrument's soft X-ray detection capability, 
restricting the usable Resolve bandpass to $\sim$1.7--12\,keV. 
In practice, this means Resolve alone cannot constrain the abundances of $\alpha$-elements lighter than Si. 
We therefore support the Resolve data with \textit{XRISM} Xtend, 
\textit{XMM-Newton} RGS and EPIC, and \textit{Chandra} ACIS. 
Since the Fe~\textsc{xxv}~He$\alpha$ and Fe~\textsc{xxvi}~Ly$\alpha$ complex used in our kinematic analysis lies at 6.6--7.0 keV, 
well above the gate-valve-closed threshold, this soft-band restriction does not affect the diagnostic lines used for our turbulence measurement.

\begin{figure*}
\centering
  \vspace*{17pt}
 \includegraphics[width=16cm]{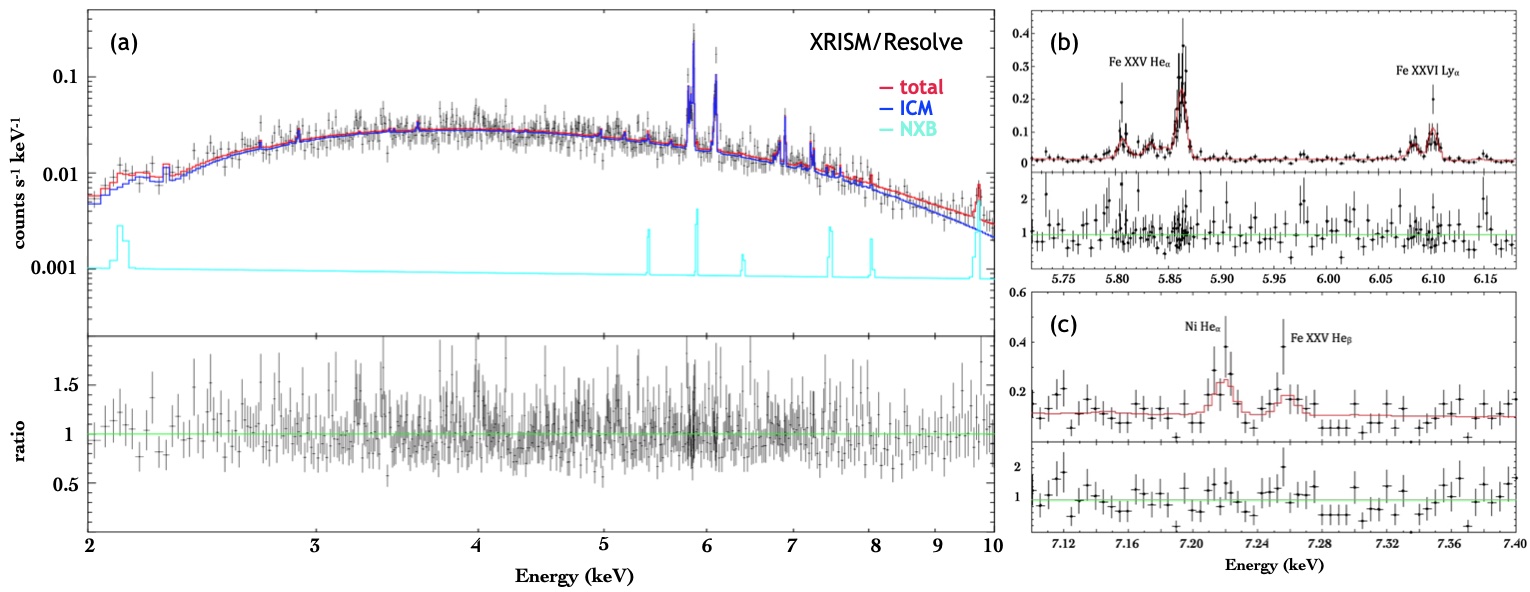}
  \caption{\textit{XRISM}/Resolve spectrum of the core of A1413 (black points with $1\sigma$ error bars) fitted with a single-temperature,
velocity-broadened \textsc{bvapec} model (red), 
shown over the full $2$--$10$\,keV fitting band (panel a) with the corresponding data-to-model ratio below. 
Panels (b) and (c) show close-up views of the Fe\,\textsc{xxv}~He$\alpha$ and Fe\,\textsc{xxvi}~Ly$\alpha$ complex ($5.7$--$6.15$\,keV) 
and the Ni\,\textsc{xxvii}~He$\alpha$ and Fe\,\textsc{xxv}~He$\beta$ complex ($6.7$--$7.0$\,keV), respectively, 
with their corresponding residual ratios shown in the lower sub-panels.}\label{spectrum_resolve}
\end{figure*}

The data were processed with the \texttt{xapipeline} tool following the standard screening criteria outlined in the \textit{XRISM} ABC Guide
v2.01.\footnote{{https://heasarc.gsfc.nasa.gov/docs/xrism/analysis/abc\_guide/xrism\_abc.html}}
We include only high-resolution primary (Hp;\texttt{ITYPE=0}) events, 
which comprise more than 99\% of the 2--10\,keV event population in the observation. 
After filtering a net exposure time of 104.65\,ks is obtained.
 
Xtend event files were reprocessed using the standard \texttt{xtdpipeline} routine. 
Hot and flickering pixels were removed using the standard pipeline screening criteria implemented in the \texttt{xtdpixclip} task, 
and the on-board calibration sources located at either edge of the Xtend FoV were masked prior to further processing. 
The cleaned and filtered event file was used to create an image in the 0.5--10\,keV band (Fig.~\ref{fig:a1413images}, 
right panel), with a net exposure of 98.22\,ks (Table~\ref{tab:obslog}).

\subsection{XMM-Newton/EPIC \& RGS}
\label{subsec:xmm}
 
We analysed four \textit{XMM-Newton} observations carried out between 2007 and 2008 
(ObsIDs 0502690101, 0502690201, 0551280101, and 0551280201; Table~\ref{tab:obslog}). 
Centred on A1413 with a small offset of 1.4 arcmin from the core, 
these observations are well suited to Reflection Grating Spectrometer (RGS) analysis. 
The MOS and PN detectors were operated with the thin filter in Full Frame and Extended Full Frame modes, respectively. 
Data reduction was performed with \textit{XMM} \textsc{SAS} v21.0, 
using the \texttt{emchain} task for MOS and \texttt{epchain} for PN.
The RGS events were then processed with \texttt{rgsproc}. 
For each instrument we extracted a background light curve in $100$ s bins from events on CCD 9.
The Gaussian-fitted mean of the distribution $\mu$ is centered near 0.02 counts s$^{-1}$ for all observations.
Good Time Interval (\textsc{gti}) files were generated with \texttt{tabgtigen} 
by retaining only intervals with a count-rate threshold of $\mu \pm 2\sigma$, which is below 0.06 counts s$^{-1}$.  
The final, flare-filtered net exposure times for MOS1, MOS2, PN, and RGS for each observation are summarised in Table~\ref{tab:obslog}.
  
\subsection{Chandra/ACIS}
\label{subsec:chandra}
 
The \textit{Chandra}/ACIS observations were carried out in March 2004 and February 2005, 
with exposure times of 76.05\,ks (ObsID 5003) and 37.15\,ks (ObsID 5002), respectively (Table~\ref{tab:obslog}). 
The cluster centre was positioned on the ACIS-I2 and ACIS-I1 chips, respectively, 
using the  \textsc{vfaint} telemetry mode (Fig.\ref{fig:a1413images} centre). 
Using \textsc{ciao} v4.18 and CalDB v4.12.4, we followed the standard data reduction procedure,
removing bad pixels and columns with the \texttt{chandra\_repro} task.
After screening, the total filtered effective exposure times were 75.05\,ks and 36.67\,ks, respectively. 
Point sources were identified with the \textsc{ciao} \texttt{wavdetect} tool in the $0.5$--$7$~keV band, 
using a false-detection probability threshold \texttt{sigthresh}~$=3.2 \times 10^{-5}$, corresponding to a one-sided $4\sigma$ Gaussian equivalent.

\begin{table*}
\caption{Best-fitting spectral parameters for the core of
A1413, from spectra extracted from a common sky region matched to the Resolve field of view. 
$N_{\rm H}$ was frozen at the Galactic value $1.82\times10^{20}$~cm$^{-2}$; 
abundances are relative to the protosolar values of \citet{Lodders2009}. 
A dash indicates that the element is not constrained by that instrument's bandpass. 
$\sigma_v$ is the line-of-sight velocity broadening, measured by Resolve alone.
Fits were performed over $2$--$10$~keV (Resolve), 0.6--10~keV (Xtend), $0.5$--$7$~keV (EPIC and ACIS) and 0.2--2.0~keV (RGS).
Uncertainties are $1\sigma$ ($68$~\%).
\label{tab:specfit}}   
\begin{center}
\setlength{\tabcolsep}{6pt}
\begin{tabular}{ l ccccc}
& \multicolumn{2}{c}{\textit{XRISM}} &  \multicolumn{2}{c}{\textit{XMM-Newton}}   & \textit{Chandra} \\ 
\hline
Parameter & {Resolve} & {Xtend} & EPIC & RGS & ACIS\\ 
\hline
$kT$ (keV) & 7.06$_{-0.43}^{+0.33}$ & 8.06$\pm{0.19}$ & 7.16$\pm0.06$ & 7.94$_{-1.61}^{+2.20}$ & 7.64$_{-0.36}^{+0.43}$ \\
O & $-$ & 0.26$_{-0.25}^{+0.66}$ & $-$ & 0.33$_{-0.23}^{+0.29}$ & $-$ \\
Ne & $-$ & 0.11$_{-0.11}^{+0.66}$ & 0.65$\pm0.18$ & 0.96$_{-0.44}^{+0.61}$ & $-$ \\
Mg & $-$   & 0.16$_{-0.16}^{+0.57}$ & 0.64$\pm0.16$ & 0.14$_{-0.14}^{+0.68}$ & 1.48$_{-0.41}^{+0.42}$  \\
Si & $-$ & 0.38$_{-0.31}^{+0.33}$ & 0.23$\pm0.09$ & 0.29$_{-0.28}^{+0.76}$ & 0.88$\pm{0.23}$ \\
S & 0.51$_{-0.40}^{+0.50}$ & 0.54$_{-0.45}^{+0.47}$ & 0.21$_{-0.14}^{+0.15}$ & $-$ & 1.11$_{-0.46}^{+0.51}$ \\
Ar & 0.48$_{-0.48}^{+0.58}$ & $-$ & $-$ & $-$ & $-$ \\
Ca & 0.34$_{-0.34}^{+0.42}$ & $-$ & $-$ & $-$ & $-$ \\
Fe & 0.38$\pm0.03$ & 0.45$\pm0.04$ & 0.38$\pm0.02$ & 0.45$_{-0.27}^{+0.59}$ & 0.32$_{-0.03}^{+0.04}$ \\
Ni & 0.45$_{-0.23}^{+0.26}$ & 0.19$_{-0.19}^{+0.45}$ & 0.71$\pm0.20$ & $-$ & $-$ \\
$\sigma_v$ (km/s) & $170^{+25}_{-24}$ & $-$ & $-$ & $-$ & $-$ \\
C-stat/dof & 16274/15988 & 1608/1572 & 3745/3892 & 2282/2130 & 557/444 \\
\hline
\end{tabular}
\end{center} 
\end{table*}

\section{Spectral Analysis}
\label{sec:spectral}
We fit the Resolve spectra using \textsc{xspec} v12.15.0 \citep{Arnaud1996} and the atomic database \textsc{AtomDB} v3.1.3 \citep{Foster2012}. 
All abundances are relative to \citet*{Lodders2009} protosolar values of \texttt{lpgs}. 
Best-fit values were determined by minimizing the C-statistic \citep{Cash1979}.
Although \textit{XRISM}/Resolve is prone to cross-pixel contamination because its half-power diameter ($1.3^{\prime}$) is 
comparable to the $3^{\prime}\times3^{\prime}$ field of view \citep{Hayashi2024}, 
our analysis of A1413 is unaffected by this localized photon leakage, 
since our spectral extraction encompasses the entire detector array without dividing the FOV into subregions.
Fig. \ref{fig:a1413images} shows the broad-band mosaic images from \textit{XMM--Newton}/EMOS in the 0.3--10 keV,  
\textit{Chandra}/ACIS in the 0.5--7 keV and \textit{XRISM}/Xtend in the 0.5--10 keV, 
overlaid with Resolve FoV and RGS extraction regions.

Source spectra were extracted by integrating high-resolution primary grade events over the FoV of Resolve. 
Pixel~12 (the out-of-aperture calibration pixel) and pixel~27, 
which exhibits anomalous gain excursions not synchronised with the $^{55}$Fe fiducial intervals, were excluded throughout.
Pixels~31 and~32 were excluded as well: a \textit{Chandra} point source falls within their projected footprint 
near the north-western edge of the array (Fig.~\ref{fig:a1413images}, centre), 
and because the Resolve half-power diameter is comparable to the array size, 
positional exclusion alone does not guarantee that its contribution is negligible. 
In Appendix~\ref{app:pointsources} we compare fits retaining the full array (\textsc{allpix}) with fits 
excluding pixel~31, pixel~32 and both pixels (WO\_31, WO\_32 and WO\_3132; Table~\ref{tab:spectral_fit_comparison}). 
Every parameter is stable to within $0.3\sigma$ across the four configurations, so the point source has no measurable effect on the results; 
we adopt WO\_3132 as the baseline for all Resolve fits quoted in this paper (Table~\ref{tab:specfit}), 
since it removes the source by construction rather than relying on that stability.
Redistribution matrix files (\textsc{rmf}s) were generated with the \texttt{rslmkrmf} task using the large (`L') size configuration, 
and were rescaled by the Hp fraction in the 2--10 keV band (excluding low-resolution secondary events) to preserve the flux normalisation.
Ancillary response files (\textsc{arf}s) were produced with \texttt{xaarfgen},
which accounts for the detector efficiencies, the effective area of the X-ray Mirror Assembly (XMA) and its point-spread function (PSF). 
We adopted an exposure-corrected \textit{Chandra} image of A1413 in the 2--8\,keV band -- 
the band most closely matching the nominal Resolve bandpass -- as the input sky brightness distribution, 
so that the extended surface-brightness profile of the cluster is correctly propagated through the response.
The non-X-ray background (\textsc{nxb}) and sky background (\textsc{cxb} plus Galactic foreground) were modelled using the templates provided by
\textsc{heasarc}\footnote{{https://heasarc.gsfc.nasa.gov/docs/xrism/proposals/nxb\_sky\_rsl\_bgd.html}}.
Spectra were binned according to the optimal binning scheme of \citet{Kaastra2016}, 
which sets the bin width from the instrumental resolution and the local counts, 
subject to a further requirement of at least one count per bin, and were fitted with the C-statistic \citep{Cash1979}.
 
 
The Xtend spectrum of the A1413 core spectrum was extracted from a rectangular region matching 
the Resolve FoV footprint on the detector (see Fig.\ref{fig:a1413images} right).
The \textsc{rmf} files were generated using the \texttt{xtdrmf} tool, while a uniform \textsc{arf} was created with \texttt{xaarfgen}: 
using an exposure-corrected \textit{Chandra}/ACIS image of A1413 as the input sky model. 
To estimate the sky background, we used \textit{XRISM}/Xtend observations of A1413. 
With its large FoV, Xtend captures a significant portion of the sky surrounding the cluster, 
extending well beyond the $R_{200}$ radius of A1413 
\citep[$\approx2.24$ Mpc $\approx14.8^\prime$;][]{Hoshino2010}.
To isolate the background, we extracted source-free spectra from a region located $15^\prime$ from the cluster center, 
safely beyond $R_{200}$ and free of detectable cluster emission. 
The \textsc{nxb} was modelled with a power-law continuum plus several Gaussian components representing instrumental fluorescence lines, fitted simultaneously with the source spectrum as an additive background component.

\begin{figure*}
\centering
  \vspace*{17pt}
 \includegraphics[width=16cm]{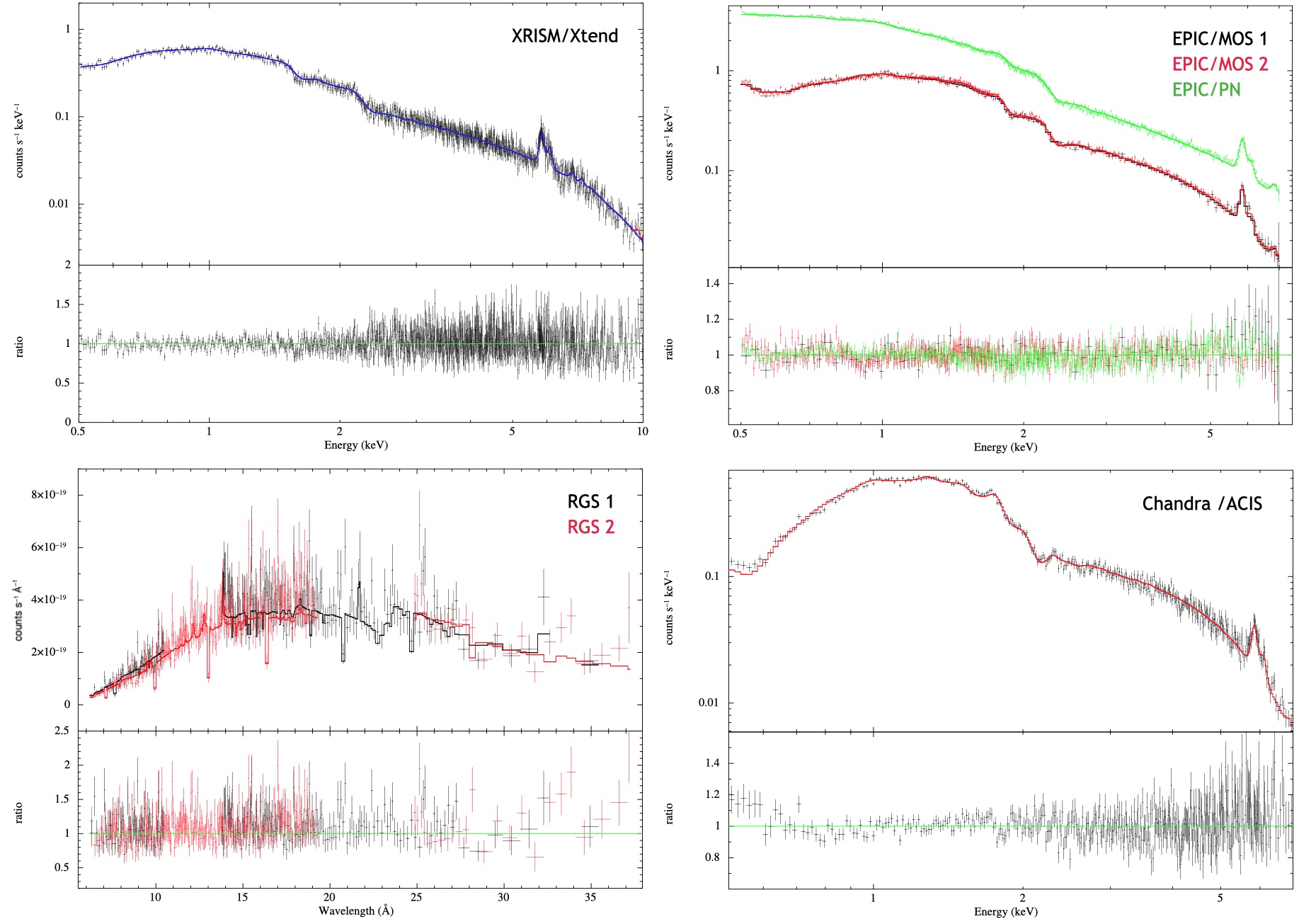}
  \caption{Spectral fits to the A1413 core for the four legacy instruments, 
  extracted from the region matched to the Resolve FoV and fitted with a \texttt{TBabs*vapec} model.  
  Clockwise from top left: \textit{XRISM}/Xtend, \textit{XMM--Newton}/EPIC (MOS1 black, MOS2 red, PN green), 
  \textit{Chandra}/ACIS, and \textit{XMM--Newton}/RGS~1 (black) and RGS~2 (red). 
  Data are shown with $1\sigma$ error bars, best-fitting models as solid lines, and data-to-model ratios below each panel. 
  The EPIC cameras were fitted simultaneously with parameters tied and normalisations free, which accounts for the offset of the PN spectrum. 
  Spectra are rebinned to a minimum significance of $4\sigma$ per bin for display only. 
  Best-fitting parameters are given in Table~\ref{tab:specfit}.}
\label{fig:legacyspectra}
\end{figure*} 

For both \textit{XMM--Newton} and \textit{Chandra}, source spectra were extracted from the same sky region matched to the Resolve FoV, 
ensuring that all four instruments sample an identical volume of the ICM (See Fig.\ref{fig:a1413images}).
The background spectra were extracted from a point-source-free region on the same chip, by avoiding the chip gaps and the CCD edges. 
Source and background spectra, together with the associated \textsc{rmf}s and \textsc{arf}s, were generated with the SAS meta-task \texttt{especget}, which runs \texttt{evselect}, \texttt{rmfgen} and \texttt{arfgen} and computes the corresponding \texttt{BACKSCAL} areas. 
The \texttt{extendedsource} option was enabled so that the \textsc{arf} is appropriate for diffuse emission filling the extraction region.
A similar approach has been adopted for cluster spectroscopy with EPIC by e.g. \citet{Takey2011}. 
The spectra of the four observations were combined separately for each camera using \texttt{epicspeccombine} tool, 
giving three merged spectra (MOS1, MOS2 and pn). 
These were not co-added further, but fitted simultaneously with the model parameters tied between cameras, 
the normalisations being left free to absorb residual cross-calibration differences in effective area.
\textit{Chandra} spectra and responses were generated with \texttt{specextract} using \texttt{weight\_rmf=yes}, 
so that both \textsc{arf} and \textsc{rmf} are weighted by the spatial distribution of counts within the extraction region
--- appropriate here given the extent of the region relative to the ACIS spectral response variations. 
The two observations were subsequently combined with \texttt{combine\_spectra},
which sums the source and background counts and produces
exposure-weighted mean response files. 
The merged EPIC and ACIS spectra were grouped with \texttt{ftgrouppha} using the optimal binning scheme of  \citet{Kaastra2016}, 
over the 0.5--7 keV band, consistent with the treatment of the Resolve data. 
For display purposes the spectra are further rebinned to a minimum significance of $4\sigma$ per bin.
 
The dispersive nature of RGS does not permit an arbitrary choice of extraction region; 
instead, the \texttt{xpsfincl} parameter of \texttt{rgsproc} selects events within a specified fraction of the telescope point spread function. 
We  adopted \texttt{xpsfincl}$=95$, corresponding to a cross-dispersion aperture of $\sim$$3\arcmin$, 
which provides good overlap with the XRISM/Resolve FoV and therefore allows a direct comparison between the instruments. 
The extraction was centred on the cluster core, 
passed to \texttt{rgsproc} via \texttt{srcstyle=radec}. 
Background spectra were constructed from events falling outside the 99 \% PSF contour (\texttt{xpsfexcl}$=99$), 
and we verified that these regions are free of contamination from the extended cluster emission. 
The associated response matrices for both instruments were generated simultaneously by \texttt{rgsproc}. In this work we consider only the first-order spectra.

\section{Thermal structure and chemical composition}
\label{sec:results}
\subsection{Is the ICM single-phase?}
\label{sec:singlephase}
 
We tested \texttt{tbabs}*(\texttt{bvapec}+\texttt{bvapec}) a 2T model, 
in which the second thermal component was allowed an independent temperature, normalisation and velocity broadening, 
while the elemental abundances were tied between the two components (the statistics are insufficient to decouple them). 
The fit returns $kT_{1} = 7.30 \pm 0.44$\,keV and $kT_{2} = 1.82 \pm 0.78$\,keV, 
with $C = 16268.2$ for $15985$ dof, compared with $C = 16274.3$ for $15988$ dof for the single-temperature model:
an improvement of $\Delta C = 6.1$ for $3$ additional free parameters. 
We do not convert this into a rejection probability: 
the normalisation of the second component lies on the boundary of the allowed parameter space, 
so the likelihood-ratio statistic does not follow the $\chi^{2}$ distribution assumed by such a conversion \citep{Protassov2002}. 
The case against the second component rests instead on its own behaviour.
 
The second component is not required by the data: $kT_{2}$ settles against the $2$\,keV lower edge of the fitting band 
and its velocity broadening, $\sigma_{v,2} = 28 \pm 503$\,km\,s$^{-1}$, is unconstrained --- 
the behaviour of a component drifting towards the boundary of the accessible parameter space rather than of a genuine detection. 
This is consistent with expectation for A1413, which is morphologically relaxed but lacks a strong cool core \citep[e.g.][]{Pratt2002, Vikhlinin2005}. 
\citet{Hoshino2010} likewise describe the $\sim$7.5\,keV gas of A1413 with a single temperature over the full XIS $0.5$--$10$\,keV band, 
though from a pointing offset from the core (Section~\ref{sec:obs}) and at a resolution that constrains multiphase structure only weakly.
We therefore adopt the single-temperature model throughout the remainder of this analysis, 
with the caveat that it describes the gas the Resolve bandpass can see: 
a phase below $\sim2$\,keV would contribute negligibly to the $2$--$10$\,keV spectrum regardless of its emission measure, 
so our adoption of a single temperature constrains the hot phase and is silent about any cooler one. 
 
The second component is not required by the data: $kT_{2}$ settles at the low-energy edge of the Resolve bandpass, 
which the closed gate valve limits to $\ge 2$\,keV, and its velocity broadening, $\sigma_{v,2} = 28 \pm 503$\,km\,s$^{-1}$, is unconstrained. 
Both are characteristic of a component drifting towards the boundary of the accessible parameter space rather than of a genuine detection.
This is consistent with expectation for A1413, which is morphologically relaxed but lacks a strong cool core \citep[e.g.][]{Pratt2002, Vikhlinin2005}, 
and with the single-temperature description of $\sim$7.5\,keV gas reported from \textit{Suzaku} results by \citet{Hoshino2010}, 
although the coarser spectral resolution of XIS provides only weak constraints on multiphase structure. 
We adopt the single-temperature model throughout the remainder of this analysis.

\subsection{Temperature measurement}
\label{sec:temperature}

The Resolve spectrum is well described by a single absorbed thermal component, giving $kT = 7.06^{+0.33}_{-0.43}$~keV 
together with a line-of-sight velocity dispersion $\sigma_v = 170^{+25}_{-24}$~km~s$^{-1}$ (C-stat/dof $=16274/15988$);
the velocity broadening is interpreted in Section~\ref{sec:kinematics}.
This is the first microcalorimeter measurement of the A1413 core, 
and the first in which the temperature is constrained by spectrally resolved Fe~K emission in addition to 
the continuum shape rather than by the continuum alone, as in CCD spectroscopy. 
The remaining instruments give $kT = 8.06\pm0.19$~keV (Xtend), $7.16\pm0.06$~keV (EPIC), $7.94^{+2.20}_{-1.61}$~keV (RGS) 
and $7.64^{+0.43}_{-0.36}$~keV (ACIS); all values are listed in Table~\ref{tab:specfit}.
Expressed in units of the combined uncertainty, Resolve, EPIC and ACIS are mutually consistent: 
Resolve agrees with EPIC to $0.3\sigma$ and with ACIS to $1.2\sigma$, and EPIC agrees with ACIS to $1.3\sigma$. 
The Xtend measurement is the single exception:  it exceeds EPIC by 0.90~keV ($4.5\sigma$) and Resolve by 1.00~keV ($2.6\sigma$), 
while remaining consistent with the less precise ACIS value ($0.9\sigma$). 
The spread across the sample is therefore driven by one instrument rather than by a systematic division between observatories. 
We emphasise that the quoted errors are statistical only; 
for EPIC in particular the $\pm0.06$~keV uncertainty is far below the level of the known effective-area systematics, 
so comparisons at this precision are governed by calibration rather than by counting statistics. 
EPIC lies 0.48~keV ($\sim$6\%) below ACIS. 
The same sense of offset was also reported by \citet{Vikhlinin2005}, 
who found the \textit{XMM--Newton} temperature of this cluster to be $\sim$10\% lower
than the \textit{Chandra} value and attributed the difference to cross-calibration. 
It also follows the systematic, temperature-dependent ACIS--EPIC offset later quantified for the HIFLUGCS sample by \citet{Schellenberger2015}. 
Our offset is smaller than in those works and is not formally significant, so we apply no cross-calibration correction.

We do not interpret the Xtend offset as a physical measurement of hotter gas. 
The Xtend and EPIC spectra were extracted from the same sky region (Section~\ref{sec:spectral}), 
so the two instruments sample an identical volume of the ICM with identical emission weighting, 
and any difference between them must be instrumental in origin. 
Because the bremsstrahlung cut-off of a $\sim7$~keV plasma lies at or beyond the upper end of the CCD bandpass, 
the fitted temperature is set by the ratio of soft-band to hard-band flux rather than by the cut-off itself, 
so a modest under-correction of the low-energy effective area hardens the apparent spectrum and biases $kT$ upwards. 
Discrepancies of this magnitude between \textit{XMM--Newton} and the \textit{XRISM} instruments are consistent with 
the current cross-calibration status of the two observatories \citep{XRISM2025c}, 
and the same mechanism underlies the \textit{Chandra}--\textit{XMM--Newton} temperature differences documented by \citet{Schellenberger2015}, 
who showed that the discrepancy originates below 2~keV while the effective-area shape at higher energies is consistent between instruments. 
The RGS temperature, $7.94^{+2.20}_{-1.61}$~keV, is formally consistent with every other measurement, 
but its bandpass (0.2--2.0~keV) excludes the Fe~K complex and lies far below the continuum cut-off; 
the agreement therefore reflects the size of the uncertainty rather than any real constraint on the temperature. 
We list it in Table~\ref{tab:specfit} for completeness but exclude it from the comparison above.

We adopt the Resolve measurement throughout this work. 
It is limited neither by the soft-band effective-area calibration that biases the CCD temperatures nor 
by the absence of Fe~K leverage that weakens the RGS constraint. 
Resolve cleanly separates the Fe\,\textsc{xxv} He$\alpha$ complex from the Fe\,\textsc{xxvi} Ly$\alpha$ line (Fig.~\ref{spectrum_resolve}b), 
so its temperature is set by an ionisation-sensitive line ratio and not by the continuum shape alone, 
and is correspondingly less sensitive to effective-area systematics. 
The extended-source \textsc{arf} generated with \texttt{xaarfgen} from the \textit{Chandra} surface-brightness distribution 
(Section~\ref{sec:spectral}) further accounts for the point-spread function of the X-ray Mirror Assembly in the response. 
The single-temperature description is justified in Section~\ref{sec:singlephase}; 
the adopted value should be understood as an emission-weighted temperature rather than that of a single physical phase.

\begin{figure*}
\centering
  \vspace*{17pt}
 \includegraphics[width=16cm]{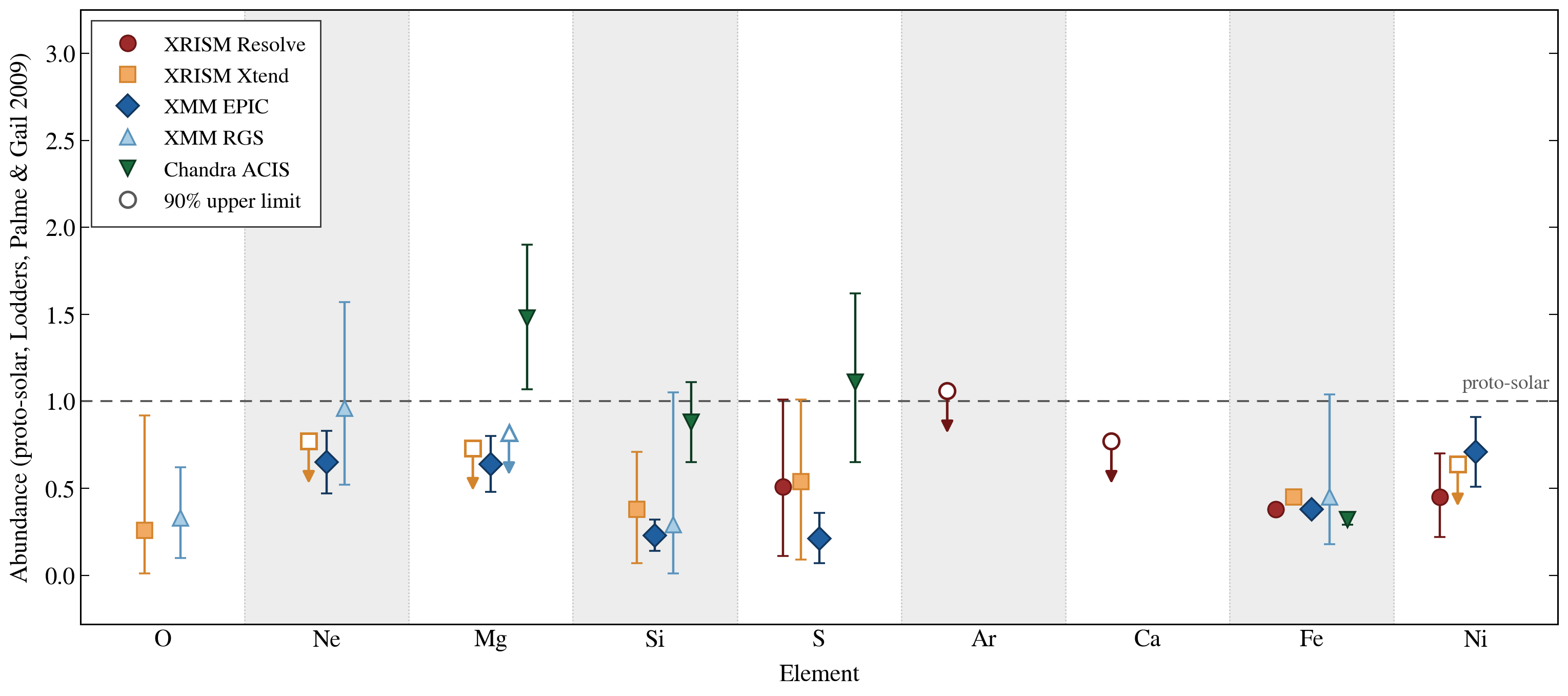}
\caption{Elemental abundances in the core of A1413, relative to the protosolar values of \citet{Lodders2009}. 
Filled symbols denote detections, open symbols with arrows upper limits, drawn wherever the lower bound reaches zero; 
all intervals are evaluated at $\Delta C = 1.0$ ($68$~\%), as in Table~\ref{tab:specfit}. 
Points are offset horizontally within each element and the dashed line marks the protosolar value. 
Missing points are elements outside an instrument's bandpass --- 
O, Ne and Mg lie below the Resolve band with the gate valve closed, Fe~K above the RGS band --- or unconstrained by the fit. 
Ar and Ca are accessible only to Resolve and are both limits ($<1.06$ and $<0.76\,Z_\odot$). 
The \textit{Chandra} Mg, Si and S values lie systematically high, as discussed in Section~\ref{sec:chandra_abundances}.}
\label{fig:abundances}
\end{figure*}

\subsection{Elemental abundances \label{sec:abundances}}
 
The five datasets considered here sample the same projected region of A1413, but differ markedly in bandpass, spectral resolution, 
and background systematics, so the abundances derived from them do not carry equal weight.  
The abundances measured by each instrument are listed in Table~\ref{tab:specfit} and shown in Fig.~\ref{fig:abundances}. 
No single instrument constrains the full set of elements, and which measurements are informative follows directly from the fitted bandpass. 
Resolve is restricted to $2$--$10$~keV and  can access only S, Ar, Ca, Fe and Ni; the O, Ne and Mg lines fall below the bandpass entirely. 
The RGS band contains O, Ne and Mg but no Fe~K, while EPIC, Xtend and ACIS span both regimes. 
The five data sets are therefore complementary rather than redundant.
 
\subsubsection{Iron}
 
Iron is the only element measured to better than $10$~\% by any instrument and provides the reference against 
which the remaining abundances must be read. 
Resolve gives Fe $= 0.38\pm0.03\,Z_\odot$, identical to the EPIC value of $0.38\pm0.02\,Z_\odot$. 
The two CCD instruments bracket this value: Xtend returns $0.45\pm0.04\,Z_\odot$ and ACIS $0.32^{+0.04}_{-0.03}\,
Z_\odot$, lying $1.4\sigma$ above and $1.2\sigma$ below the Resolve measurement respectively and differing from one another by $2.3\sigma$.
The Resolve, EPIC and RGS values agree to within $0.3\sigma$. 
The residual scatter, though not formally significant, is carried entirely by the two CCD
imagers: a weighted mean of all five gives $0.379 \pm 0.014\,Z_\odot$ with $\chi^2 = 6.0$ for 4 dof (p=0.20), 
of which Xtend and ACIS contribute $6.0$. 
The spread between the extremes, ACIS ($0.32$) and Xtend ($0.45$), 
is comparable to the effective-area cross-calibration differences reported for these instruments \citep{Schellenberger2015}.
Adopting a corresponding systematic term, we quote ${\rm Fe} = 0.38 \pm 0.03\,({\rm stat}) \pm 0.04\,({\rm sys})\ Z_\odot$.
The RGS value, $0.45^{+0.59}_{-0.27}\,Z_\odot$, is consistent with all of these but too weakly constrained to discriminate between them.
Xtend estimates both the highest temperature ($kT = 8.06 \pm 0.19$~keV) and the highest iron abundance (Fe $= 0.45 \pm 0.04\,Z_\odot$); 
we regard these offsets as a systematic effect rather than as independent results. 
The Xtend spectrum was extracted from a region matching the Resolve FoV, so differences in the physical aperture cannot account for them. 
To isolate the bandpass dependence we refitted the Xtend spectrum over the $2$--$10$~keV Resolve band, 
freezing O, Ne, Mg and Si at their broad-band values since their lines fall outside this range. 
Both parameters move towards the Resolve values: 
$kT = 7.48^{+0.28}_{-0.26}$~keV and Fe $= 0.42^{+0.04}_{-0.03}\,Z_\odot$, for C-stat/dof $=1321/1326$. 
The offsets from Resolve fall from $2.6\sigma$ to $1.0\sigma$ in temperature and from $1.4\sigma$ to $0.9\sigma$ in iron. 
The temperature shift alone, $0.58$~keV, is three times the broad-band statistical uncertainty, 
so the discrepancy is generated almost entirely below $2$~keV and not by the Fe~K complex or by the continuum within the Resolve band.

This is the behaviour expected if the low-energy effective area is slightly under-corrected. 
The bremsstrahlung cut-off of a $\sim7$~keV plasma lies at or beyond the upper end of the CCD bandpass, 
so the fitted temperature is set by the ratio of soft-band to hard-band flux rather than by the cut-off itself;
a hardened soft band therefore biases $kT$ upwards, and because the Fe~\textsc{xxv} He$\alpha$ emissivity falls with temperature in this regime,
the fit compensates with a higher Fe abundance. 
The two offsets are thus a single effect rather than two independent ones, 
which is consistent with their moving together and in the same proportion when the soft band is removed.
A residual $0.9$--$1.0\sigma$ difference remains, consistent with zero; the band restriction therefore accounts for the discrepancy.
In the restricted fit neither S nor Ni is constrained --- both peg at the hard limit of zero, with 1$\sigma$ upper limits of $0.42$ 
and $0.78\,Z_\odot$ ---confirming that the Xtend values for these elements carry no weight against the Resolve measurements.
We therefore exclude the Xtend abundances from the combination and retain Xtend only for the broad-band continuum characterisation.
 
We adopt Fe$=0.38\,Z_\odot$ as the reference value: it is derived from the resolved Fe--K He$\alpha$ complex and 
is therefore free of the Fe--L atomic-data and modelling degeneracies that affect CCD measurements \citep{Gu2019}, 
and it coincides with the independent EPIC value.

\subsubsection{Elements accessible to Resolve}

Resolve is the only instrument to constrain Ar and Ca, and the only one to measure Ni free of the Fe~K$\beta$ blend. 
At CCD resolution the Ni~K$\alpha$ line at $7.5$~keV is unresolved from Fe~K$\beta$, 
so the two abundances are strongly covariant and the recovered Ni depends on the assumed Fe~K$\beta$ emissivity; 
Resolve separates these features cleanly (Fig.~\ref{spectrum_resolve}c). 
We measure Ni $= 0.45^{+0.26}_{-0.23}\,Z_\odot$, a $2.0\sigma$ detection, corresponding to Ni/Fe~$=1.18\pm0.64$ and consistent with solar. 
The EPIC value, $0.71\pm0.20\,Z_\odot$ (Ni/Fe~$=1.9\pm0.6$), is higher by $0.8\sigma$; 
the offset is not significant, but its direction is what the Fe~K$\beta$ blend would produce. 
Resolve therefore measures Ni free of the Fe--K$\beta$ blend that unavoidably affects CCD determinations, 
and free of the instrumental Ni\,K$\alpha$ fluorescence line in EPIC-pn. 
Although the EPIC ratio is nominally the more precise, that precision rests on flux which Resolve resolves as belonging to Fe; 
we therefore regard the Resolve measurement as the more accurate and adopt it throughout.

Sulphur is constrained to $0.51^{+0.50}_{-0.40}\,Z_\odot$ ($1.3\sigma$), consistent with the Xtend ($0.54^{+0.47}_{-0.45}$) and 
EPIC ($0.21^{+0.15}_{-0.14}$) values at $0.05\sigma$ and $0.7\sigma$ respectively. 
Argon and calcium are not significantly detected. 
Stepping the fit statistic over each in turn gives smooth, single-minimum curves with best-fitting values of $0.48$ and $0.34\,Z_\odot$; 
setting either abundance to zero degrades the fit by only $\Delta C = 2.7$ and $2.3$, corresponding to $1.7\sigma$ and $1.5\sigma$. 
We therefore quote them as upper limits, Ar~$<1.06\,Z_\odot$ and Ca~$<0.76\,Z_\odot$, 
evaluated at the $\Delta C = 1.0$ ($1\sigma$) criterion used for all intervals in Table~\ref{tab:specfit}.

\subsubsection{Light elements}

The elements below Si are constrained principally by EPIC, which measures 
Ne~$=0.65\pm0.18$, Mg~$=0.64\pm0.16$ and Si~$=0.23\pm0.09\,Z_\odot$ (all $\geq2.5\sigma$). 
The RGS independently detects Ne at $0.96^{+0.61}_{-0.44}\,Z_\odot$ ($2.2\sigma$), 
consistent with EPIC at $0.7\sigma$, and provides the only oxygen constraint, $0.33^{+0.29}_{-0.23}\,Z_\odot$ (O/Fe~$=0.7$).
The corresponding EPIC ratios are Ne/Fe~$=1.7$, Mg/Fe~$=1.7$ and Si/Fe~$=0.6$.

\subsubsection{The \textit{Chandra} abundance pattern}
\label{sec:chandra_abundances} 

The ACIS $\alpha$-element abundances are systematically and significantly higher than those of every other instrument: 
Mg~$=1.48^{+0.42}_{-0.41}$, Si~$=0.88\pm0.23$ and S~$=1.11^{+0.51}_{-0.46}\,Z_\odot$, 
exceeding the EPIC values by $1.9\sigma$, $2.6\sigma$ and $1.9\sigma$ respectively. 
Combined with the lowest Fe abundance in the sample, these give Mg/Fe~$\simeq4.6$, Si/Fe~$\simeq2.8$ and S/Fe~$\simeq3.5$ times solar.
Ratios of this magnitude are not produced by any physically motivated combination of SNIa and SNcc yields, 
and exceed by a large factor the near-solar values measured in the cores of other massive clusters \citep{Mernier2017, Simionescu2019}. 
Because the anomaly affects all three $\alpha$-elements coherently while Fe remains within $1.3\sigma$ of the EPIC value, 
it cannot be attributed to the Fe normalisation alone. 
The ACIS fit also has the poorest statistic in the sample (C-stat/dof $=557/444= 1.25$, against $0.96$--$1.07$ for the others) and shows coherent residuals below $\sim1$~keV (Fig.~\ref{fig:legacyspectra}), where the ACIS contamination correction is least certain and where the Mg and Si lines lie. 
We therefore attribute the pattern to a residual calibration issue in the ACIS soft-band response and exclude the ACIS abundances from the enrichment
analysis, retaining them in Table~\ref{tab:specfit} for completeness.
 
\subsubsection{Systematic uncertainties}
 
Three effects merit comment. First, all fits assume a single thermal component; 
fitting multiphase gas with a 1T model biases the recovered Fe abundance low through the shape of the Fe-L complex
\citep{Gastaldello2021, Sarkar2022}. 
This mechanism operates below $\sim1.2$~keV, so it affects the CCD measurements but not the Resolve fit,
which is driven by Fe~K; the single-temperature description is justified
in Section~\ref{sec:singlephase}.
 
Second, resonant scattering in the optically thick Fe~\textsc{xxv} He$\alpha$ $w$ line can suppress its flux and bias Fe low, 
reaching a factor of $\sim1.3$ within the central $\sim30$~kpc of Perseus \citep{Hitomi2018}. 
We tested this directly by excluding a $40$~eV band centred on the redshifted $w$ centroid ($5.863$~keV) 
while retaining the optically thin $y$ and $z$ lines: Fe rises from $0.381$ to $0.386\,Z_\odot$, a $0.1\sigma$ change. 
Removing the entire He$\alpha$ complex increases it only to $0.403\,Z_\odot$, still within the $90$~\% interval of the full fit. 
The shift has the sign expected from resonant scattering but is negligible against the statistical error, 
as anticipated for the lower central density and $\sim450$~kpc FoV of A1413.
 
Third, all abundances are quoted relative to the protosolar values of \citet{Lodders2009}; 
comparisons with published measurements adopting a different reference require rescaling.


\begin{table}
\caption{Screened abundance ratios relative to iron in the core of A1413.
Column~3 lists the instruments combined for each element; 
Xtend is excluded throughout and ACIS is excluded for Mg, Si and S. 
Uncertainties are $1\sigma$, as in Table~\ref{tab:specfit}; 
column~4 gives the deviation from the protosolar value \citep{Lodders2009} in units of $\sigma$.}
\label{tab:xfe}
\begin{tabular}{lccr}
\hline
Ratio & Value & Instruments & Deviation $(\sigma)$ \\
\hline
O/Fe 	& $0.73\pm0.91$ & RGS           	& $-0.3$ \\
Ne/Fe 	& $1.73\pm0.47$ & EPIC, RGS     	& $+1.6$ \\
Mg/Fe 	& $1.45\pm0.39$ & EPIC, RGS     	& $+1.2$ \\
Si/Fe 	& $0.61\pm0.24$ & EPIC, RGS     	& $-1.6$ \\
S/Fe  	& $0.63\pm0.36$ & Resolve, EPIC 	& $-1.0$ \\
Ar/Fe 	& $1.26\pm1.40$ & Resolve       	& $+0.2$ \\
Ca/Fe 	& $0.89\pm1.00$ & Resolve       	& $-0.1$ \\
Ni/Fe 	& $1.18\pm0.64$ & Resolve       	& $+0.3$ \\
\hline
\multicolumn{4}{c}{$\chi^{2} = 7.7$ for 8 d.o.f. against a solar pattern ($p = 0.47$)} \\ \hline
\end{tabular}
\end{table}

\subsection{Enrichment by SN explosions}
 
 The eight ratios of Table~\ref{tab:xfe} do not carry equal weight. 
 O/Fe, Ar/Fe and Ca/Fe have fractional uncertainties in excess of $100$~per cent and sit within a small fraction of their errors of solar; 
 they constrain no enrichment history and we do not interpret their central values. 
 Ni/Fe is better measured ($54$~per cent fractional uncertainty) but likewise consistent with solar, 
 and is discussed separately below because it probes a different question. 
The four remaining ratios --- Ne/Fe, Mg/Fe, Si/Fe and S/Fe --- depart from solar by $1.0$--$1.6$, 
and together account for $7.4$ of the total $\chi^{2}=7.7$ by which the pattern differs from solar.

The lighter $\alpha$-elements O, Ne and Mg are synthesised in the hydrostatic C- and Ne-burning shells of massive stars 
and ejected essentially unaltered by the core-collapse explosion, with a negligible SNIa contribution \citep[e.g.][]{Nomoto2013}; 
their yields scale steeply with progenitor mass,making them the cleanest tracers of the core-collapse contribution. 
We measure Ne/Fe~$=1.73\pm0.47$ and Mg/Fe~$=1.45\pm0.39$, both above solar in the same sense. 
Among the intermediate-mass elements, Si/Fe~$=0.61\pm0.24$ and S/Fe~$=0.63\pm0.36$ are both sub-solar and agree closely, 
indicating a consistent deficit rather than an anomaly.
 
 Taken as a physical signature these deviations do not survive scrutiny,  because they pull in opposite directions. 
 A super-solar Ne/Fe requires a reduced SNIa fraction, whereas a sub-solar Si/Fe requires an enhanced one; 
 no single value of $R({\rm Ia})$ accommodates both, since raising the core-collapse contribution to reproduce the neon ratio deepens the silicon deficit. The two are therefore mutually inconsistent as physical effects. 
 Their common feature is instrumental: Ne/Fe, Mg/Fe and Si/Fe are all measured by EPIC and the RGS below $\sim2$~keV,
the band in which we identified residual calibration differences between instruments in Section~\ref{sec:abundances}, 
and which lies outside the range accessible to Resolve with the gate valve closed. 
At $kT\simeq7$~keV the Fe--L complex is essentially absent and the soft-band lines are intrinsically weak, 
so the continuum placement underlying the Ne\,\textsc{x} measurement is the least well constrained in our pattern. 
These three ratios alone contribute $\chi^{2} = 6.4$ for 3~dof ($p = 0.09$), $83$~per cent of the total, 
whereas the four ratios measured by Resolve (S, Ar, Ca and Ni relative to Fe) give 
$\chi^{2} = 1.2$ for 4~dof ($p = 0.88$) and are fully consistent with a solar pattern. 
The formal rejection of the solar pattern in Table~\ref{tab:xfe} is thus driven entirely by the soft band, 
and we attribute the departures to soft-band systematics rather than to enrichment history.

The Ni/Fe ratio is of a different character and does not bear on $R({\rm Ia})$.
Nickel production is governed by electron capture during high-density burning,
so Ni/Fe discriminates between SNIa explosion channels rather than between supernova types. 
Progenitors near the Chandrasekhar mass ($M_{\rm Ch}$) yield enhanced $^{58}$Ni, 
whereas sub-$M_{\rm Ch}$ detonations characteristically underproduce it \citep[e.g.][]{Seitenzahl2013,LeungNomoto2018}. 
Our value, Ni/Fe~$=1.18\pm0.64$, is measured by Resolve free of the Fe~K$\beta$ blend,
and is therefore the ratio least affected by the systematics discussed above.
Even so, it lies within $0.3\sigma$ of solar, and its $1\sigma$ interval, $0.54$--$1.82$, spans the predictions of both explosion channels. 
The present data therefore do not discriminate between near- and sub-$M_{\rm Ch}$ progenitors; 
a factor of two improvement in precision would be required to do so.

 
   \begin{figure}
  \includegraphics[width=\columnwidth]{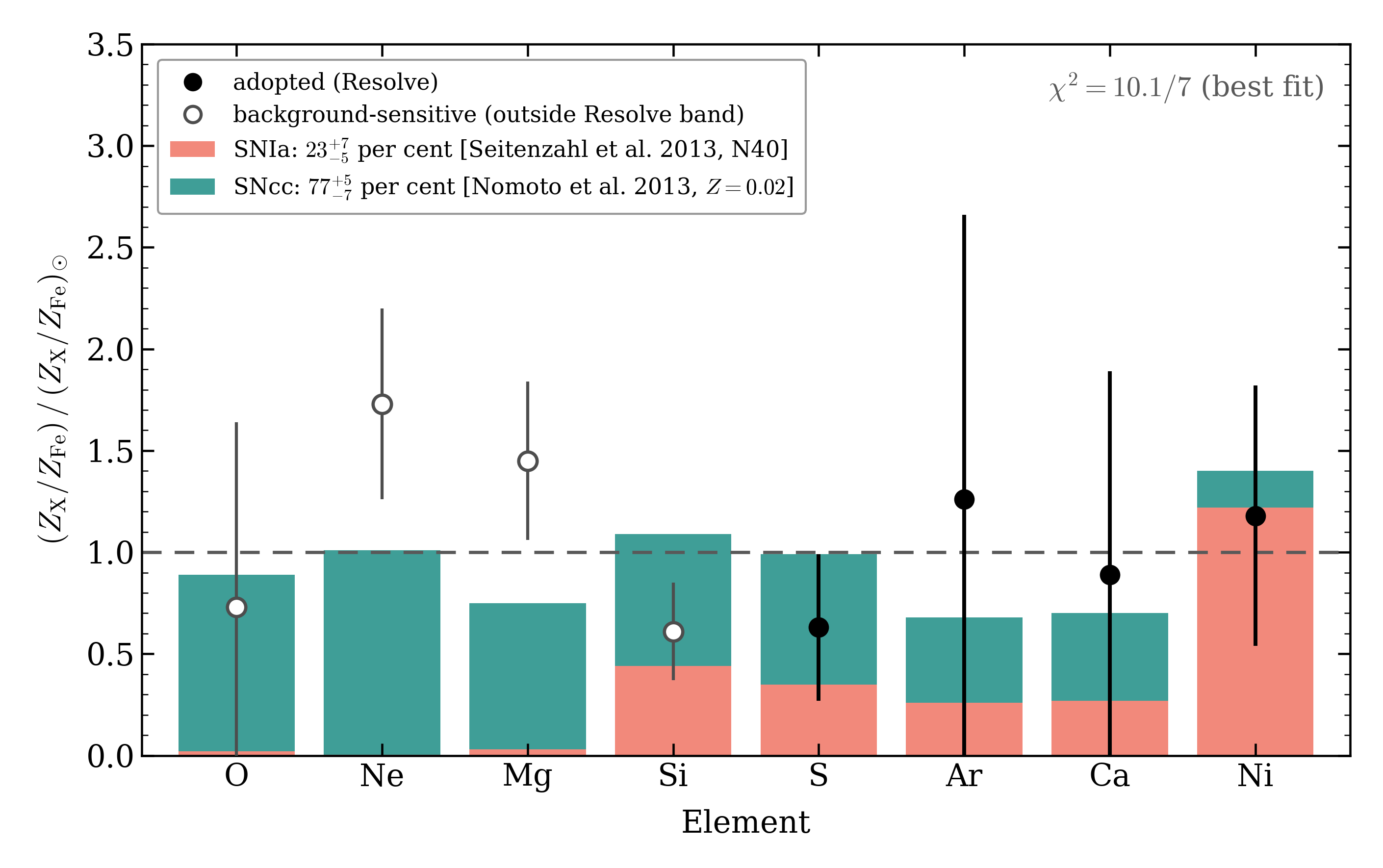}
  \caption{X/Fe abundance ratios in the core of A1413, relative to protosolar \citep{Lodders2009}, 
  with the best-fitting SNIa $+$ SNcc yield decomposition stacked bars; models and fitted fractions as labelled). 
  Open circles mark the background-sensitive soft-band ratios (O, Ne, Mg and Si), 
  which lie outside the band accessible to Resolve with the gate valve closed. 
  Error bars are $1\sigma$, as in Table~\ref{tab:xfe}. Fe is the normalising element and is omitted. 
  The dashed line indicates the solar value.}  \label{fig:xfe}
\end{figure} 

\subsubsection{Supernova yield fitting}
\label{sec:sneratio}
Subject to these caveats, we examine what enrichment history the measured pattern would imply. 
To decompose the enrichment of the A1413 core into its Type~Ia and core-collapse contributions, 
we fit the measured abundance pattern with the open-source \textsc{SNeRatio} code \citep{Erdim2021}, 
which compares the observed X/Fe ratios to a number-weighted linear combination of tabulated nucleosynthesis yields 
and returns the SNIa fraction of the total enriching population, $R({\rm Ia})={\rm SNIa}/({\rm SNIa}+{\rm SNcc})$.
Core-collapse yields are averaged over a Salpeter IMF \citep{Salpeter1955} between $10$ and $50$~M$_{\odot}$; 
SNIa yields require no such integration.
Because the model is normalised to Fe, the fit constrains only the \emph{shape} of the pattern and is independent of the absolute Fe abundance. 
All ratios are quoted relative to the protosolar values of \citet{Lodders2009}. 
Since $R({\rm cc}) = 1 - R({\rm Ia})$, the SNIa fraction is the single free parameter of the fit. 
All $\chi^{2}$ values in this subsection are evaluated with the $1\sigma$ uncertainties of Table~\ref{tab:xfe}. 
They remain approximate in one respect: the ratios share a common Fe denominator and so are not strictly independent.

We fit the eight ratios O, Ne, Mg, Si, S, Ar, Ca and Ni relative to Fe, exploring near-$M_{\rm Ch}$ delayed detonations \citep{Seitenzahl2013},
near-$M_{\rm Ch}$ deflagrations \citep{Fink2014} and sub-$M_{\rm Ch}$ detonations \citep{Shen2018,LeungNomoto2020} 
in combination with the SNcc yields of \citet{Nomoto2013} for $Z_{\rm init}=0$--$0.05$. 
The best-fitting combination is the N40 delayed-detonation model of \citet{Seitenzahl2013} together 
with the $Z_{\rm init}=0.02$ SNcc yields of \citet{Nomoto2013}, 
giving $\chi^{2}/\nu = 10.1/7$ and $R({\rm Ia}) = 23^{+7}_{-5}$~per cent (Fig.~\ref{fig:xfe}), 
hence $R({\rm cc}) = 77^{+5}_{-7}$~per cent.  
On a straightforward reading, the A1413 core would then be core-collapse dominated in its enrichment.
Three considerations argue against that conclusion.

First, the decomposition is not statistically required: a single-component model reproduces the measured ratios with comparable goodness of fit.
A purely solar abundance pattern --- which has no free parameters at all --- gives $\chi^{2} = 7.7$ for 8 degrees of freedom, 
roughly a quarter \emph{lower} than the best-fitting yield combination achieves with one free parameter and one degree of freedom fewer. 
Because a solar pattern is not a member of the SNIa\,$+$\,SNcc mixture family for any $R({\rm Ia})$, 
the two models are non-nested and this ordering is not a contradiction; 
it is also independent of the error normalisation, since both statistics scale by the same factor. 
None of the yield sets we explored improves on the assumption that the core is enriched in solar proportions. 
A comparable result was found for the Perseus core, 
where a protosolar enrichment pattern describes the high-resolution measurements well  ($\chi^{2} = 10.7$ for 10~$\nu$) 
yet remains challenging to reproduce with linear combinations of existing supernova yield calculations \citep{Simionescu2019}. 
Our finding is thus consistent with a known limitation of current yield models rather than specific to A1413.

Second, the constraint rests almost entirely on the ratios we have just argued are unreliable. 
Restricting the fit to the four ratios measured directly by Resolve (S, Ar, Ca and Ni relative to Fe) 
gives $\chi^{2}/\nu = 1.9/3$ and $R({\rm Ia}) = 35^{+65}_{-20}$~per cent. 
The central value is consistent with the eight-ratio result at the $0.6\sigma$ level, but the confidence interval widens from 
$12$ to $85$ percentage points --- a sevenfold inflation --- and its upper end reaches the physical boundary $R({\rm Ia}) = 1$, 
so that these four ratios on their own cannot exclude a purely SNIa-enriched population. 
The low $\chi^{2}_{\nu} = 0.6$ reflects the size of the uncertainties on Ar/Fe and Ca/Fe rather than the quality of the model. 
This behaviour is expected, because the intermediate-mass elements Si, S, Ar and Ca receive contributions 
from both SNcc and SNIa in broadly comparable amounts \citep{Nomoto2013,Mernier2016}, 
so their X/Fe ratios vary far less steeply with $R({\rm Ia})$ than those of O, Ne and Mg, for which the SNIa contribution is negligible; 
in addition, Ar/Fe and Ca/Fe carry almost no statistical weight. 
Removing the three soft-band ratios thus removes essentially all of the leverage, 
and we do not consider the SNIa fraction constrained by the present data.

Third, the residuals of the best-fitting model are element-specific rather than random:  it overpredicts Si/Fe and S/Fe while underpredicting Mg/Fe. 
This is the same qualitative mismatch reported for other cool cores, 
and in particular the systematic overproduction of S/Ar by current yield models \citep{Simionescu2019,Mernier2025}; 
our Ar/Fe and Ca/Fe uncertainties ($\sim100$~per cent) are far too large to contribute to that discussion, 
which for the Perseus core rests on ratios constrained to better than $10$~per cent \citep{Simionescu2019}. 
For the same reason our data cannot test the Ca excess reported in A2029 \citep{Sarkar2025}, 
for which Ca/Fe would need to be measured to a precision several times better than achieved here.
 
We conclude that the XRISM core spectrum of A1413 is consistent with enrichment in solar proportions, 
that the departures from that pattern are confined to the soft band and are most plausibly instrumental, 
and that the SNIa fraction is not usefully constrained by these data: 
the ratios with the leverage to constrain it are the ones we cannot trust, and those we can trust ---
$R({\rm Ia}) = 35^{+65}_{-20}$~per cent --- leave the answer open between $15$ and $100$~per cent. 
A measurement of Si/Fe and S/Fe within the Resolve band, 
which the removal of the gate valve will provide, is what this decomposition requires.

\section{Gas kinematics and non-thermal pressure support}
\label{sec:kinematics}
 
The Resolve spectrum constrains both the centroid and the width of the Fe\,\textsc{xxv}~He$\alpha$ and Fe\,\textsc{xxvi}~Ly$\alpha$ complexes,
and so yields a direct measurement of the line-of-sight (LOS) velocity field of the intracluster medium (ICM). 
Here we convert the best-fitting redshift and velocity-broadening parameters of the single-temperature \textsc{bvapec} model adopted in
Sect.~\ref{sec:results} (Table~\ref{tab:specfit}) into bulk and turbulent velocities. 
Throughout we adopt $\gamma = 5/3$ and a mean molecular weight $\mu = 0.61$, appropriate for a fully ionised plasma at the metallicity measured here. 
All velocities are corrected to the Solar-system barycentre.
 
\subsection{Bulk velocity}
\label{sec:bulk}
 
The meaningful bulk motion is that of the X-ray gas relative to the collisionless component of the potential, 
traced by the brightest cluster galaxy (BCG). Following \citet{XRISM2025d},

\begin{equation}
v_{\rm bulk} = c\,\frac{z_{\rm ICM}-z_{\rm BCG}}{1+z_{\rm BCG}}.
\label{eq:vbulk}
\end{equation}
 
\noindent
With $z_{\rm ICM} = 0.142880 \pm 0.000080$ (statistical) from the Resolve fit and $z_{\rm BCG} = 0.142972$ \citep{Fouque1992}, 
the redshift difference is $\Delta z = -9.2\times10^{-5}$, 
carrying $\pm 8.0\times10^{-5}$ from the Resolve fit and $\pm 24\times10^{-5}$ from the optical redshift (see below), and

\begin{equation}
v_{\rm bulk} = -24 \pm 26\,(z_{\rm ICM}) \pm 71\,(z_{\rm BCG})\ {\rm km\,s^{-1}}.
\label{eq:vbulkval}
\end{equation}
 
\noindent
The first term combines the statistical uncertainty on $z_{\rm ICM}$ ($21$~km\,s$^{-1}$) with 
the $\approx0.3$~eV in-orbit energy-scale accuracy of Resolve at $5.4$--$9.0$~keV \citep{Eckart2025,Porter2025}, 
which is $15$~km\,s$^{-1}$ at the observed energy of the Fe\,\textsc{xxv}~He$\alpha$ complex; 
the second is the uncertainty on the optical velocity of the BCG. 
\citet{Fouque1992} quote no formal error; 
\citet{Castagne2012} give $42\,832 \pm 71$~km\,s$^{-1}$ for the same galaxy from the Catalogue of Principal Galaxies, 
and we adopt that uncertainty here. It is conservative in that we do not apply the $(1+z)^{-1}$ factor of Eq.~\ref{eq:vbulk}, 
which would reduce it to $62$~km\,s$^{-1}$. 

We adopt the instrumental term directly without a calibration-pixel check for this observation; 
published per-observation values are $0.32$~eV for A2029 \citep{XRISM2025a} 
and $0.30$--$0.51$~eV for A2319 \citep{XRISM2025e}, and even the upper end is negligible against the optical term. 
Adding the two terms of Eq.~\ref{eq:vbulkval} in quadrature gives $\sigma_{\rm tot} = 76$~km\,s$^{-1}$: 
the core gas is blueshifted with respect to the BCG at only $0.3\sigma$, 
and the $90$~per cent confidence interval is $-149 < v_{\rm bulk} < +101$~km\,s$^{-1}$. 
The corresponding bulk Mach number is $\mathcal{M}_{\rm bulk} = -0.018 \pm 0.056$. 
We stress that this uncertainty enters only the line centroid; the velocity dispersion of Section~\ref{sec:turb}, 
and hence the non-thermal pressure fraction and mass bias, are unaffected.

The central value is not sensitive to the choice of optical redshift. 
The three published velocities for the A1413 BCG --- 
$42\,862$~km\,s$^{-1}$ \citep{Fouque1992}, $42\,844 \pm 100$~km\,s$^{-1}$ \citep[Humason et al. 1956,
as quoted by][]{Castagne2012} and $42\,832 \pm 71$~km\,s$^{-1}$ \citep[PGC, {\it ibid.}]{Castagne2012} --- 
give $v_{\rm bulk} = -24$, $-9$ and $+2$~km\,s$^{-1}$ respectively against our $cz_{\rm ICM} = 42\,834$~km\,s$^{-1}$.
All three place the ICM at rest with respect to the BCG to within $25$~km\,s$^{-1}$; 
what limits the measurement is the width of the optical error bar, not the value adopted.

The gas in the A1413 core is therefore at rest with respect to the BCG. This is not evidence against sloshing.
Deep \textit{Chandra} imaging reveals a surface-brightness edge $\sim$$400$ kpc north of the centre \citep{Vikhlinin2005},
the morphological signature of a sloshing cold front \citep{Markevitch2007}, 
so cold gas is being displaced in this cluster independently of anything the kinematics show. 
What Resolve constrains is therefore not whether A1413 is sloshing, but the orientation of the plane in which it does so.

Simulations of minor mergers predict a line-of-sight velocity difference of $\mathcal{M} = 0.2$--$0.3$ 
when that plane is oriented along the line of sight \citep{ZuHone2016}, 
which for the sound speed of A1413 ($c_{s} = 1359$~km\,s$^{-1}$; Eq.~\ref{eq:cs}) corresponds to $270$--$410$~km\,s$^{-1}$ --- 
comparable to the velocity differences XRISM resolves across the cores of A2319 and A3667 \citep{XRISM2025e,Omiya2026}, 
and a factor $1.8$--$2.8$ above our $90$~per cent upper limit of $149$~km\,s$^{-1}$. 
A sloshing plane containing the line of sight is therefore excluded. 
Combined with the cold front, which establishes that sloshing is occurring, 
the null bulk velocity becomes a geometric constraint rather than a null result: 
the sloshing plane of A1413 must lie close to the plane of the sky --- the configuration inferred for A2029 \citep{PaternoMahler2013,XRISM2025d}. 
We note that the edge itself lies beyond the $226$~kpc half-width of the Resolve field, 
so this constraint applies to the sloshing motion of the core rather than to the gas at the front.
A dedicated spectroscopic redshift of the A1413 BCG, which would reduce the dominant term in our error budget by an order of magnitude, 
is what a stronger statement requires.
 
The core is also displaced from the cluster as a whole. 
\citet{Castagne2012} measure a mean cluster velocity of $42\,387 \pm 111$~km\,s$^{-1}$ and a galaxy velocity dispersion of
$1196 \pm 83$~km\,s$^{-1}$ from $256$ members, placing the BCG $416 \pm 115$~km\,s$^{-1}$ from 
the cluster rest frame for our adopted $z_{\rm BCG}$ (and $390 \pm 116$~km\,s$^{-1}$ if the PGC velocity is used instead), 
or about a third of the galaxy velocity dispersion. The ICM tracks
the BCG rather than the cluster mean, so the core and its central galaxy form
a kinematically coherent system moving within the cluster potential --- the
configuration also found in A2199 \citep{Suda2026}.

\subsection{Turbulent velocity and Mach number}
\label{sec:turb}
 
The \textsc{bvapec} velocity-broadening parameter, fitted independently of the thermal Doppler width and the Resolve line-spread function, 
gives $\sigma_{v} = 170^{+25}_{-24}$~km\,s$^{-1}$. 
Assuming isotropic motions, $v_{\rm 3D} = \sqrt{3}\,\sigma_{v} = 294^{+43}_{-42}$~km\,s$^{-1}$, 
while the sound speed follows from $kT = 7.06^{+0.33}_{-0.43}$~keV from the same single-temperature Resolve fit as
 
\begin{equation}
c_{s} = \sqrt{\frac{\gamma\,k_{\rm B}T}{\mu m_{p}}}
      = 1359^{+31}_{-42}\ {\rm km\,s^{-1}},
\label{eq:cs}
\end{equation}

\noindent
giving $\mathcal{M}_{\rm 1D} = 0.125^{+0.019}_{-0.018}$ and $\mathcal{M}_{\rm 3D} = 0.217^{+0.033}_{-0.031}$. 
Including the bulk term through the effective velocity of \citet{Hitomi2018}, 
$v_{\rm 3D,eff} = (3\sigma_{v}^{2}+v_{\rm bulk}^{2})^{1/2} = 295^{+43}_{-42}$~km\,s$^{-1}$, 
 raises the Mach number only from $\mathcal{M}_{\rm 3D} = 0.2166$ to $\mathcal{M}_{\rm 3D,eff} = 0.2173$, 
since the non-thermal pressure scales as the square of the velocity and the bulk term is an order of magnitude smaller than $\sigma_{v}$. 
The motions are strongly subsonic.

Two effects make $\sigma_{v}$ formally an upper limit on the turbulent component: 
it is an emission-measure-weighted average over the field of view and along the full line of sight, so any unresolved velocity gradient inflates it; 
and the Resolve point spread function (half-power diameter $\approx1.3$~arcmin) mixes emission across the array, 
so the measurement characterises the central $\sim100$~kpc (the half-power radius of $0\farcm65$) rather than a sharply bounded region.
\begin{table}
\centering
\caption{Kinematic properties of the A1413 core from XRISM/Resolve. Uncertainties are $1\sigma$ statistical unless noted; note that the optical redshift of the BCG is a factor $\sim3$ less precise than the Resolve measurement of the ICM.}
\label{tab:kinematics}
\begin{tabular}{lc}
\hline
Quantity & Value \\
\hline
$kT$ (keV)                                       		& $7.06^{+0.33}_{-0.43}$ \\
$z_{\rm ICM}$                                    	& $0.14288 \pm 0.00008$ \\
$z_{\rm BCG}$ (optical)$^{a}$              	& $0.14297 \pm 0.00024$ \\
$\sigma_{v}$ (km\,s$^{-1}$)                      	& $170^{+25}_{-24}$ \\
$v_{\rm bulk}$ (km\,s$^{-1}$)$^{b}$           & $-24 \pm 26 \pm 71$ \\
$c_{s}$ (km\,s$^{-1}$)                           	& $1359^{+31}_{-42}$ \\
$v_{\rm 3D}=\sqrt{3}\sigma_{v}$ (km\,s$^{-1}$)   & $294^{+43}_{-42}$ \\
$v_{\rm 3D,eff}$ (km\,s$^{-1}$)          	& $295^{+43}_{-42}$ \\
$\mathcal{M}_{\rm 1D}$                           	& $0.125^{+0.019}_{-0.018}$ \\
$\mathcal{M}_{\rm 3D}$                           	& $0.217^{+0.033}_{-0.031}$ \\
$\mathcal{M}_{\rm bulk}$                         	& $-0.018 \pm 0.056$ \\
$P_{\rm NT}/P_{\rm tot}$ (\%)              	& $2.5 \pm 0.7$ \\
\hline
\end{tabular}
\begin{flushleft}
\footnotesize
$^{a}$ \citet{Fouque1992}. That catalogue quotes no formal uncertainty; we adopt $\pm71$~km\,s$^{-1}$ ($\delta z = 2.4\times10^{-4}$), 
the error on the Catalogue of Principal Galaxies velocity of the same galaxy given by \citet{Castagne2012}.
$^{b}$ First term: uncertainty on $z_{\rm ICM}$ ($21$~km\,s$^{-1}$ statistical, combined with the $\pm0.3$~eV Resolve energy-scale systematic,
$15$~km\,s$^{-1}$; Sect.~\ref{sec:bulk}). Second term: uncertainty on $z_{\rm BCG}$ (footnote~$a$). Combined in quadrature, $76$~km\,s$^{-1}$.
\end{flushleft}
\end{table}
\subsection{Non-thermal pressure and hydrostatic mass}
\label{sec:pnt}

Isotropic motions contribute a non-thermal pressure $P_{\rm NT} = \frac{1}{3}\rho_{\rm gas}v_{\rm 3D}^{2} = \rho_{\rm gas}\sigma_{v}^{2}$ \citep{Lau2009,Eckert2019}, which as a fraction of the total pressure depends on the Mach number alone, 
\begin{equation}
\alpha \equiv \frac{P_{\rm NT}}{P_{\rm tot}}
 = \frac{\mathcal{M}_{\rm 3D}^{2}}{\mathcal{M}_{\rm 3D}^{2} + 3/\gamma}
 = 2.5 \pm 0.7\ {\rm per\ cent}.
\label{eq:alpha}
\end{equation}
\noindent
Using $\mathcal{M}_{\rm 3D,eff}$ in place of $\mathcal{M}_{\rm 3D}$ changes this from $2.54$ to $2.56$~per cent, i.e.\ well within the uncertainty. 
Two effects make this a bound rather than a measurement: $\sigma_{v}$ is an emission-measure-weighted average that any unresolved velocity gradient
inflates (Section~\ref{sec:turb}), and it is averaged over a region substantially larger than the radius within which the cooling time is measured
(Section~\ref{sec:scale}). We therefore read $2.5 \pm 0.7$~per cent as an upper bound on the non-thermal support of the core.
Table~\ref{tab:kinematics} collects the derived quantities and 
Table~\ref{tab:comparison} places A1413 among cluster cores observed at high spectral resolution. 
Our value is indistinguishable from that of the relaxed cluster A2029 ($2.6\pm0.3$ \%; \citealt{XRISM2025a}),
higher than the quiescent cores of Ophiuchus and A2199 ($1.4\pm0.2$ \%; \citealt{Fujita2025,Suda2026}), 
and lower than cores hosting powerful AGN such as Perseus ($3.9\pm0.8$ \%; \citealt{Hitomi2016,XRISM2026}) 
and Hydra\,A ($4.5\pm0.5$ \%; \citealt{Rose2025}). 
A1413 thus reinforces the emerging result that non-thermal support in cluster cores lies in a narrow $1$--$5$ \% range, 
at the low end of the values predicted by cosmological simulations \citep{Nelson2014,Vazza2018,Angelinelli2020}.

Since A1413 is a standard calibrator for hydrostatic masses
\citep[e.g.][]{Vikhlinin2006,Pratt2019}, this has a direct practical
consequence. Following \citet{Ettori2022}, the mass bias is
\begin{equation}
b(r) = \frac{\alpha(r) + A(r)}{1 + A(r)}, \qquad
A(r) = P_{\rm tot}\,\frac{{\rm d}\alpha}{{\rm d}r}
       \left(\frac{{\rm d}P_{\rm T}}{{\rm d}r}\right)^{-1};
\end{equation}
a single pointing cannot constrain the gradient, but adopting $A \ll \alpha$ as found for relaxed cores \citep{XRISM2025d} gives $b \la 2.5$~per cent. Turbulence is therefore not a viable explanation for the $\sim20$--$30$ \% bias invoked to reconcile 
X-ray with lensing and SZ masses \citep{vonderLinden2014,Hoekstra2015,Planck2016} ---
at least at the small radii probed here. 
Two caveats bound this statement. 
Simulations predict $\alpha$ rising to $10$--$20$ \% by $R_{200}$ \citep{Nelson2014,Angelinelli2020}, 
precisely the radii a central pointing cannot reach, 
so our result anchors the inner boundary of $\alpha(r)$ rather than constraining $b$ globally. 
And if the velocity field is preferentially in the plane of the sky, 
$v_{\rm 3D}$ could be underestimated by up to a factor $\sim2$ \citep{ZuHone2018}, 
raising $\alpha$ to $\sim10$ \% --- still well below what the mass tension requires.

\begin{table}
\centering
\caption{Non-thermal pressure fraction in cluster cores observed at high
spectral resolution, ordered by the extent $l$ of the region sampled.}
\label{tab:comparison}
\setlength{\tabcolsep}{4pt}
\begin{tabular}{lcccc}
\hline
Cluster & $l$ & $\sigma_{v}$ & $\sigma_{v}^{\rm A1413}(l)^{a}$ & $P_{\rm NT}/P_{\rm tot}$ \\
        & (kpc) & (km\,s$^{-1}$) & (km\,s$^{-1}$) & (\%) \\
\hline
\textbf{A1413} (this work) 	& $452$ 	& $170^{+25}_{-24}$ 	& --    	& $2.5\pm0.7$ \\
A2029 (core)               	& $263$ 	& $169\pm10$        		& $142$ 	& $2.6\pm0.3$ \\
Hydra\,A                   		& $189$ 	& --                			& --    	& $4.5\pm0.5$ \\
A3571 (core)               	& $138$ 	& $118^{+14}_{-15}$ 	& $115$ 	& $1.3\pm0.3$ \\
A2199                      		& $106$ 	& $100\pm5$         		& $105$ 	& $1.4\pm0.2$ \\
Perseus                    		& $65$  	& $164\pm10$        		& $89$  	& $3.9\pm0.8$ \\
Ophiuchus (inner)          	& $53$  	& $115\pm7$         		& $83$  	& $1.4\pm0.2$ \\
Centaurus                  	& $42$  	& $117^{+9}_{-9}$   		& --    	& $3.3\pm0.6$ \\
\hline
\end{tabular}
\begin{flushleft}
\footnotesize
$^{a}$ Our $\sigma_{v}$ rescaled from $452$~kpc as $\sigma \propto l^{1/3}$
(Section~\ref{sec:scale}); illustrative only, and blank where the aperture is
not established.

References: A2029 -- \citet{XRISM2025a}; Hydra\,A -- \citet{Rose2025};
A3571 -- \citet{McCall2026}; A2199 -- \citet{Suda2026}; Perseus --
\citet{Hitomi2016,XRISM2026}; Ophiuchus -- \citet{Fujita2025}; Centaurus --
\citet{XRISM2025b}. The A2029, Hydra\,A, Perseus and Centaurus values are as
compiled by \citet{Fujita2025}.
\end{flushleft}
\end{table}
 
\subsection{The scale of the measurement}
\label{sec:scale}

Table~\ref{tab:comparison} compares dispersions measured over very different physical volumes. 
The $3\arcmin\times3\arcmin$ Resolve field subtends $452$~kpc at $z = 0.143$, against $263$~kpc for A2029, 
$138$~kpc for A3571 and $65$~kpc for Perseus, so $\sigma_{v}$ measured here is an emission-weighted average 
over a region $1.7$--$7$ times larger in linear extent than in the nearby cores with which it is compared. 
The weighting is concentrated towards the centre by the surface-brightness profile, 
but the aperture is set by the array, not by the point spread function. 
\citet{McCall2026} make this point quantitatively for A3571 using the effective line-of-sight length $l_{\rm eff}$ of \citet{Zhuravleva2012}, 
showing that AGN-active cool cores are sampled on scales $\la 50$~kpc while relaxed and 
merging systems are sampled on scales of hundreds of kpc, and 
that kinematic comparisons between clusters are meaningful only at matched $l_{\rm eff}$.

This bears directly on the interpretation of our result. 
Taking a Kolmogorov cascade, $\sigma \propto l^{1/3}$, and rescaling the A3571 core measurement 
($\sigma_{v} \simeq 110$--$118$~km\,s$^{-1}$ over $l_{\rm eff} = 79^{+29}_{-22}$~kpc) to 
the $452$~kpc extent of the Resolve field gives $197$--$211$~km\,s$^{-1}$, 
above our measured $170^{+25}_{-24}$~km\,s$^{-1}$ by $1.1$--$1.6\sigma$. 
The aperture is, however, an upper bound on the scale actually sampled: 
the emission weighting is concentrated towards the centre by the surface-brightness profile, 
so the appropriate $l_{\rm eff}$ is smaller and the extrapolated dispersion lower.
Rescaling instead to the $\sim$$100$~kpc half-power radius of the Resolve PSF,
a lower bound on the sampled scale, gives $119$--$128$~km\,s$^{-1}$. 
Our measurement lies between these limits. 
The factor of $\sim1.4$ by which A1413 exceeds A3571 is thus consistent with 
the difference in sampled scale alone and does not require a higher turbulent amplitude: 
the two weak cool cores agree once the scale of the measurement is accounted for. 
We stress that this is a consistency check rather than a matched comparison, which would require
$l_{\rm eff}$ to be computed for our extraction region from the deprojected \textit{Chandra} density profile, as \citet{McCall2026} do for A3571.

The same rescaling can be applied to the other entries of Table~\ref{tab:comparison}. 
Reduced to the $138$~kpc sampled in A3571 our dispersion becomes $115$~km\,s$^{-1}$, 
within $0.2\sigma$ of the $118^{+14}_{-15}$~km\,s$^{-1}$ measured there, and 
reduced to the $109$~kpc of A2199 it becomes $106$~km\,s$^{-1}$ against $\simeq100$~km\,s$^{-1}$: 
at matched scale A1413 sits with the quiescent cores of the sample. 
Reduced to the $263$~kpc of A2029 it becomes $142$~km\,s$^{-1}$, below the $169\pm10$ measured there, 
though only by $1.2\sigma$; the apparent equality of the raw A1413 and A2029 dispersions is 
thus a coincidence of apertures differing by a factor $1.7$ in linear extent rather than a physical match. 
The cores with a recent AGN outburst history behave differently: rescaled to the $65$~kpc Hitomi 
Perseus field our dispersion is $89$~km\,s$^{-1}$ against $164\pm10$, to the $53$~kpc Ophiuchus field 
$83$~km\,s$^{-1}$ against $115\pm7$, and to the $42$~kpc Centaurus field $77$~km\,s$^{-1}$ against $\simeq120$~km\,s$^{-1}$.
Scale therefore accounts for the A1413--A2029 coincidence but not for the spread of Table~\ref{tab:comparison} as a whole: 
at matched sampling scale the AGN-active cores remain the more turbulent, as expected if feedback-driven motions dominate there. 
We stress that a single Kolmogorov scaling applied across clusters differing in temperature, 
cooling state and AGN power is illustrative rather than quantitative.

The same argument makes our non-thermal pressure fraction conservative. 
Since $\sigma_{v}$ grows with the scale over which it is averaged, 
the value measured across $452$~kpc bounds the dispersion of the innermost gas from above, 
and the $2.5 \pm 0.7$~per cent of Section~\ref{sec:pnt} is correspondingly an upper limit on 
$P_{\rm NT}/P_{\rm tot}$ within the radius where $t_{\rm cool}$ was measured. 
The conclusion that turbulence at these radii cannot supply the $20$--$30$~per cent mass bias is 
therefore strengthened, not weakened, by the size of the region sampled. 
 
\section{Summary and conclusions}
\label{sec:conclusions}
 
We have presented the first \textit{XRISM} observation of the core of A1413, 
combining $104.7$~ks of Resolve and $98.2$~ks of Xtend data with archival \textit{XMM--Newton} (EPIC, RGS) and 
\textit{Chandra} (ACIS) spectra extracted from a matched sky region. Our conclusions are as follows.
 
\begin{enumerate}
\item The Resolve spectrum is well described by a single velocity-broadened thermal component with $kT = 7.06^{+0.33}_{-0.43}$~keV. 
A second component is not required ($\Delta C = 6.1$ for 3 extra parameters), and we find no evidence for
the multiphase structure seen in comparably massive cool-core systems.

\item Comparison across the five instruments reveals systematic offsets of $10$--$20$ \% in temperature and abundance. 
Refitting the Xtend spectrum over the $2$--$10$~keV Resolve band lowers $kT$ from $8.06$ to $7.48$~keV and 
Fe from $0.45$ to $0.42\,Z_\odot$, localising the discrepancy below $2$~keV and identifying it as instrumental rather than physical; 
the \textit{Chandra} $\alpha$-element excess has the same origin.

\item We measure Fe~$= 0.38 \pm 0.03\,({\rm stat}) \pm 0.04\,({\rm sys})\,Z_\odot$ from the resolved Fe--K He$\alpha$ complex, 
free of the Fe--L modelling degeneracies affecting CCD spectroscopy and in exact agreement with the independent EPIC value. 
Excluding the optically thick $w$ line changes it by $1.3$\%, so resonant scattering does not measurably bias this measurement. 
Separating Ni~K$\alpha$ from Fe~K$\beta$ for the first time in this cluster gives Ni/Fe consistent with solar; 
Resolve also provides the only constraints on Ar and Ca, both upper limits.
 
\item The screened abundance pattern is close to solar. 
The residual departures --- super-solar Ne/Fe and sub-solar Si/Fe --- 
pull in opposite directions on the SNIa fraction and cannot be reconciled by any single enrichment history. 
A supernova-yield fit to the eight X/Fe ratios returns $R({\rm Ia}) = 23^{+7}_{-5}$~per cent, 
whereas the four ratios Resolve measures directly give $R({\rm Ia}) = 35^{+65}_{-20}$~per cent ($\chi^{2} = 1.9$ for $3$ d.o.f.), 
with the upper bound running into the $R({\rm Ia}) = 100$ \% ceiling. 
The two determinations are formally consistent, but only because the hard-band value is unconstrained: 
the lines Resolve resolves permit essentially the full physically allowed range. 
The apparent precision of the eight-ratio result is therefore carried entirely by the soft-band ratios, 
and we do not regard the SNIa fraction as constrained by the present data.
 
\item The core gas is at rest with respect to the BCG,
$v_{\rm bulk} = -24 \pm 26\,(z_{\rm ICM}) \pm 71\,(z_{\rm BCG})$~km~s$^{-1}$, disfavouring a coherent line-of-sight sloshing flow. 
The uncertainty is dominated by the optical redshift of the BCG rather than by Resolve, 
and a modern spectroscopic redshift of that galaxy would tighten it by an order of magnitude. 
Resolving the Fe\,\textsc{xxv}~He$\alpha$ and Fe\,\textsc{xxvi}~Ly$\alpha$ complexes gives $\sigma_{v} = 170^{+25}_{-24}$~km~s$^{-1}$, 
or $\mathcal{M}_{\rm 3D} = 0.217^{+0.033}_{-0.031}$, independent of that limitation.
 
\item The corresponding non-thermal pressure fraction is $2.5 \pm 0.7$~per cent,
formally an upper bound because $\sigma_{v}$ is averaged over a region larger
than the cooling radius, implying a hydrostatic mass bias $b \la 2.5$~per cent
within the core. Rescaled to a common sampling scale,  
A1413 matches the other weak cool core A3571 to $0.2\sigma$ and lies marginally below the strong cool-core cluster A2029, 
despite lacking a strong cool core: the absence of central cooling does not translate into elevated turbulence.
Turbulence at these radii cannot account for the $20$--$30$ \% bias invoked to reconcile X-ray with lensing and SZ masses.
 
\end{enumerate}
 
Taken together, these results show that the limiting factor on abundance and temperature measurements in massive cluster cores is no longer photon
statistics but cross-instrument calibration below $2$~keV --- precisely the band Resolve cannot access with the gate valve closed. 
Gate-valve-open observations, extending the microcalorimeter bandpass to the O, Ne and Mg lines, would remove the principal systematic identified here and convert the supernova budget of A1413 from an open question into a measurement.


\section*{Acknowledgements}
 
We gratefully acknowledge financial support from the Scientific and Technological Research Council of T\"urkiye 
(T\"{U}B\.{I}TAK) under project number 126F221. The \textit{XRISM} observation of A1413 analysed here
(ObsID 201049010) was obtained under a Guest Observer programme led by H.~Akamatsu, 
and was retrieved from the public archive following the expiry of its exclusive-use period. 
We thank the \textit{XRISM} team for the operation of the observatory and for the calibration products and analysis software that made this work
possible. 

This paper is based on observations obtained with \textit{XRISM}, a joint
JAXA/NASA mission with participation by ESA; on archival observations obtained
with \textit{XMM--Newton} and the \textit{Chandra} X-ray Observatory.

\section*{Data Availability}
 
All observational data underlying this article are held in public archives.
The reduced spectra, response files and spectral-fit products generated for this analysis are available from the
corresponding author on reasonable request.

 
\appendix

  \begin{table*}
	\centering
	\caption{Comparison of \texttt{TBabs*bvapec} spectral fits to the Resolve FoV of A1413, with $N_{\rm H}$ frozen at the Galactic value ($1.82\times10^{20}\,{\rm cm^{-2}}$). \textsc{allpix} denotes the fit using all array pixels other than the calibration pixel~12 and the gain-unstable pixel~27, which are excluded from every fit; WO\_31, WO\_32, and WO\_3132 denote fits in which pixel~31, pixel~32, and both pixels~31 and~32 are additionally excluded, in order to test for bias introduced by the point source coincident with these pixels. WO\_3132 is the configuration adopted in the main text.}
	\label{tab:spectral_fit_comparison}
	\begin{tabular}{lcccc}
 
	\hline
	Parameter & \textsc{allpix} & WO\_31 & WO\_32 & WO\_3132 \\
	\hline
	$kT$ (keV)          & $7.03 _{-0.38}^{+0.31}$        & $7.02 _{-0.41}^{+0.32}$        & $7.07 _{-0.42}^{+0.32}$        & $7.06 _{-0.43}^{+0.33}$        \\
	S             & $0.46 _{-0.38}^{+0.48}$        & $0.49 _{-0.39}^{+0.49}$        & $0.48 _{-0.39}^{+0.49}$        & $0.51 _{-0.40}^{+0.50}$        \\
	Ar           & $0.48 _{-0.47}^{+0.56}$        & $0.44 _{-0.44}^{+0.56}$        & $0.52 _{-0.48}^{+0.57}$        & $0.48 _{-0.48}^{+0.58}$        \\
	Ca           & $0.30 _{-0.30}^{+0.41}$        & $0.33 _{-0.33}^{+0.41}$        & $0.32 _{-0.32}^{+0.42}$        & $0.34 _{-0.34}^{+0.42}$        \\
	Fe           & $0.379\pm{0.03}$      & $0.379\pm{0.03}$      & $0.381\pm{0.03}$      & $0.382\pm{0.03}$     \\
	Ni             & $0.45 _{-0.26}^{+0.23}$        & $0.43 _{-0.23}^{+0.26}$        & $0.47 _{-0.23}^{+0.26}$        & $0.45 _{-0.23}^{+0.26}$        \\
	Redshift             & $0.142860 _{-0.00008}^{+0.00009}$ & $0.142865 _{-0.00008}^{+0.00009}$ & $0.142875 _{-0.00009}^{+0.00008}$ & $0.142880 _{-0.00008}^{+0.00008}$ \\
	Velocity (km s$^{-1}$) & $171 _{-23}^{+25}$          & $171 _{-23}^{+25}$           & $170 _{-24}^{+25}$           & $170 _{-24}^{+25}$           \\
	norm                 & $0.02219\pm{0.00101}$              & $0.02212\pm{0.00106}$              & $0.02204\pm{0.00105}$              & $0.02197\pm{0.00106}$              \\
	C-stat / dof        & $16349.5 / 15988$      & $16317.8 / 15988$      & $16309.1 / 15988$      & $16274.3 / 15988$      \\
	\hline
	\end{tabular}
\end{table*}

\section{Point-source contamination of the Resolve spectrum}
\label{app:pointsources}
 
 Before addressing the detailed structure of the ICM, we first assess the
X-ray point source lying within the projected footprint of Resolve pixels~31
and~32, close to the north-western edge of the array, at R.A.,
Dec.$_{\rm J2000}$ = $11^{\rm h}55^{\rm m}24\fs2$,
$+23\degr24\arcmin42\farcs9$ (see \textit{Chandra} image at
Fig.~\ref{fig:a1413images}, centre). Because the Resolve half-power diameter
of $1\farcm3$ \citep{Hayashi2024} is comparable to the array size, the
emission of such a source is not confined to the pixels on which it falls,
and a simple positional exclusion does not guarantee that its contribution is
negligible.

We therefore repeated the spectral fitting with these pixels excluded
individually and in combination, and compare the resulting parameters against
the full-array (\textsc{allpix}) reference. Table~\ref{tab:spectral_fit_comparison}
summarises the best-fitting \texttt{TBabs*bvapec} parameters for each subset.

Every parameter is stable across the four configurations. The temperature
varies by only $\sim$$0.05$\,keV, roughly an eighth of the
$^{+0.33}_{-0.43}$\,keV statistical uncertainty on any one fit; the iron
abundance moves from $0.379$ to $0.382\,Z_\odot$, about $0.1\sigma$; and the
velocity broadening changes by $1$\,km\,s$^{-1}$, from $171$ to
$170$\,km\,s$^{-1}$. The remaining abundances are mutually consistent but far
more loosely constrained, with uncertainties of $50$--$80$\% of the
fitted value, and carry correspondingly little weight in this comparison. Iron
is the best-determined abundance ($\sim$$8$\%) and is therefore the
most sensitive test available: its stability is the clearest indication that
the point source does not contaminate the Fe--K measurement on which our
abundance and kinematic results rest.

The one parameter to show a systematic trend is the redshift, which increases
monotonically from $0.142860$ to $0.142880$ as pixels are removed. The shift
corresponds to $5$\,km\,s$^{-1}$, a quarter of the statistical uncertainty on
$z_{\rm ICM}$, and is the largest excursion of any parameter in
Table~\ref{tab:spectral_fit_comparison}. Because it is monotonic rather than
random, it may reflect a mild velocity gradient across the array, with
pixels~31 and~32 sampling a slightly different part of the velocity field; it
is equally consistent with statistical noise, and the present data cannot
distinguish between the two. In either case the amplitude is negligible for
the bulk velocity of Section~\ref{sec:bulk}, whose error budget is dominated
by the optical redshift of the BCG. The ICM parameters are therefore
insensitive to the exclusion of pixels~31 and~32 at the $\le0.3\sigma$ level,
and the point source has no measurable effect on our \texttt{bvapec} results.

Although the checks show no bias, we adopt WO\_3132 as the baseline for all
Resolve results quoted in the main text (Table~\ref{tab:specfit}), since it is
the configuration in which the contaminating source is removed by construction
rather than shown to be harmless a posteriori. The \textsc{allpix} fit is
retained here only as the reference against which the exclusions are judged.
 
 
\bsp
\label{lastpage}

\end{document}